\documentclass[11pt,a4paper]{article}
\IfFileExists{lmodern.sty}{\usepackage[T1]{fontenc}\usepackage{lmodern}}{}
\usepackage[utf8]{inputenc}
\usepackage[left=28mm,right=28mm,top=25mm,bottom=25mm]{geometry}
\usepackage{amsmath,amssymb,mathtools}
\usepackage{amsthm}

\usepackage{array}
\usepackage{xcolor}
\usepackage[colorlinks=true,linkcolor=blue,citecolor=blue,urlcolor=blue]{hyperref}

\numberwithin{equation}{section}
\makeatletter
\renewcommand\section{\@startsection{section}{1}{\z@}%
  {-3.5ex \@plus -1ex \@minus -.2ex}%
  {2.3ex \@plus .2ex}{\normalfont\large\bfseries}}
\makeatother

\newcommand{\papertitle}{Twisting Spinning Particles\\[1mm] with Superghost Backgrounds}
\newcommand{\paperauthors}{Erik Fink$^{a}$, Jingqian Gou$^{b}$, Niklas Holbach$^{b}$, Ond\v{r}ej Hul\'{i}k$^{a,c}$}

\newcommand{\paperaffiliation}{%
  $^a$Institute for Mathematics, Ruprecht-Karls-Universit\"at Heidelberg,\\
  69120 Heidelberg, Germany\\[1mm]
  $^b$Institute for Theoretical Physics, Ruprecht-Karls-Universit\"at Heidelberg,
  Philosophenweg 19,\\
  69120 Heidelberg, Germany \\ [1mm]
  $^c$Department of Theoretical Physics and Astrophysics, Faculty of Science,
  Masaryk University,\\
  Kotl\'a\v{r}sk\'a 2, Brno, Czechia  }

\newcommand{\paperemail}{efink@mathi.uni-heidelberg.de, gou@thphys.uni-heidelberg.de, \\holbach@thphys.uni-heidelberg.de, ondra.hulik@gmail.com}

\hypersetup{%
  pdftitle={Twisting Spinning Particles with Superghost Backgrounds},
  pdfauthor={Erik Fink, Jingqian Gou, Niklas Holbach, Ond\v{r}ej Hul\'{i}k}}

\begin{document}
\hypersetup{pageanchor=false}
\begin{titlepage}
\thispagestyle{empty}
\setlength{\parindent}{0pt}

{\small\scshape \hfill \today\par}
\vspace{3mm}
\hrule height 0.5pt
\vspace{17mm}

{\raggedright\fontsize{20}{25}\selectfont\bfseries \papertitle\par}
\vspace{13mm}

{\normalsize\bfseries \paperauthors\par}
\vspace{3mm}
{\small\itshape\leftskip=1em \paperaffiliation\par}
\vspace{2mm}
{\small\leftskip=1em \textit{E-mail:}\enspace
  \begin{tabular}[t]{@{}>{\ttfamily}l@{}}\paperemail\end{tabular}\par}
\vspace{12mm}

\textbf{Abstract:}\enspace
We construct twists of spinning particles in the classical BV
formalism by giving constant backgrounds to the bosonic ghosts
of local worldline supersymmetry, in analogy with Costello and
Li's twist of supergravity, which gives a nonzero background
to the ghost of local spacetime supersymmetry.
We represent the superghosts on polynomial states and gauge selected
worldline R symmetries, imposing the shifted R constraints absolutely
and the unshifted ones relatively.
On flat targets, the $\mathcal N=2$ particle, its $\mathcal N=4$
extension on a complex target and an $\mathcal N=8$ particle with two
fermion flavors give the de~Rham, Dolbeault and polynomial linear BCOV
complexes. The last two are the linearized field content of the open
and closed B-model target-space theories.
Pairings and interactions require target-space input beyond the
reduced worldline differentials.
\par
\vspace{7mm}
{\raggedright\textbf{Keywords:}\enspace spinning particles, BV formalism,
topological twisting, holomorphic Chern--Simons, BCOV theory\par}

\vfill
\end{titlepage}
\hypersetup{pageanchor=true}

\tableofcontents
\bigskip
\hrule
\bigskip

\section{Introduction}
\label{sec:introduction}
Topological strings give a geometric description of selected sectors
of supersymmetric string theory~\cite{Witten:1988xj,Witten:1991zz}. In
the B-model, open-string field theory on a Calabi--Yau threefold leads
to holomorphic Chern--Simons theory~\cite{Witten:1992fb}, while the
closed sector is described by Kodaira--Spencer theory and its BCOV
extension~\cite{Bershadsky:1993cx,Costello:2012cy}. Their linear
complexes involve Dolbeault forms and polyvector-valued forms, with
differentials $\bar\partial$ and $\bar\partial+t\partial_\Omega$,
respectively. In this paper we use the worldline as a model for a different route to these complexes, namely twisting the complete gauged worldsheet theory of the physical string rather than an auxiliary matter CFT. Such a twist would be a worldsheet counterpart of the twist of supergravity proposed by Costello and Li~\cite{Costello:2016}.

The usual worldsheet construction starts from a matter theory with global $\mathcal N=(2,2)$ supersymmetry. The twist shifts the generator of worldsheet rotations by a $U(1)$ R charge, turning one of the supercharges into a scalar differential $Q_{\mathrm{top}}$.
The twisted energy-momentum tensor is $Q_{\mathrm{top}}$-exact,
\[
 Q_{\mathrm{top}}^2=0,
 \qquad
 T_{\mathrm{tw}}=\{Q_{\mathrm{top}},G\},
\]
where $G$ is the remaining supercurrent, which has spin two after the
twist. An analogous relation holds in the other chiral sector. This
defines the twisted matter theory on a fixed surface. Coupling it to
topological gravity gives a topological string. The fermionic partner of the worldsheet metric is then a symmetric tensor, the twisted counterpart of the vector-spinor gravitino. After gauge fixing, the measure on the moduli space of Riemann surfaces involves insertions of $G$ and its right-moving counterpart~\cite{Witten:1991zz,Bershadsky:1993cx}.

In this construction the twist acts on an auxiliary $\mathcal N=(2,2)$ matter CFT, which is then coupled to topological gravity. The physical RNS string is organized differently. Its
worldsheet supersymmetry is gauged by $\mathcal N=(1,1)$ supergravity,
and the gravitini and supersymmetry ghosts are part of the
first-quantized gauge theory. The $\mathcal N=1$ superconformal algebra
has no $U(1)$ R current, so the matter twist above requires the
enhanced $\mathcal N=(2,2)$ symmetry of special backgrounds. Even on such backgrounds, twisting the matter currents alone leaves open how the worldsheet supergravity fields, ghosts and gauge symmetries would reorganize into those of topological gravity, and how gauge fixing would produce the corresponding measure on moduli space. Topological reformulations of physical superstrings are known,
notably those of Berkovits and Vafa~\cite{Berkovits:1994vy}. We ask whether one can instead twist the full gauged RNS theory directly in the BV formalism and follow the effect of the twist through gauge fixing to the resulting state complex.

Costello and Li give a BV formulation of twisting for
supergravity~\cite{Costello:2016}. Since local spacetime supersymmetry
is a gauge symmetry, its ghost $\Gamma$ is a bosonic spinor. To define the twist, one expands the master action about a background $\Phi_0$ in which the supersymmetry ghost $\Gamma$ takes a nonzero value $\Gamma_0$. In terms of the fluctuations $\widetilde\Phi$, the twisted action and its BV differential are
\[
 \Gamma=\Gamma_0+\widetilde\Gamma,
 \qquad
 S_{\mathrm{tw}}[\widetilde\Phi]
 =S_{\mathrm{BV}}[\Phi_0+\widetilde\Phi],
 \qquad
 s_{\mathrm{tw}}=(S_{\mathrm{tw}},-)_{\mathrm{BV}}.
\]
Here $\Phi_0$ is a solution of the equations of motion of the full BV
action. For a flux-free background with constant scalars, these equations require the metric to be Ricci-flat and the ghost $\Gamma_0$ to be covariantly constant and satisfy the square-zero condition $[\Gamma_0,\Gamma_0]=0$ in the supersymmetry algebra~\cite{Costello:2016,Elliott:2020ecf}. The expansion changes the BV differential $s_{\mathrm{BV}}$ by the action
of the chosen supersymmetry, together with its ghost and antifield
completion. This realizes the twist within the gauge theory itself.

Costello and Li's proposal concerns target-space supergravity. In particular, they conjecture that BCOV theory describes a holomorphic twist of ten-dimensional type IIB supergravity on a Calabi--Yau fivefold. At the string level, the corresponding conjecture identifies the twisted type IIB string with the topological B-model~\cite{Costello:2016}. This is a
different setting from the familiar Calabi--Yau threefold topological
string, and the target-space construction does not itself supply the
corresponding operation on the worldsheet. Local worldsheet supersymmetry is a gauge symmetry of the RNS string, so its BV action contains bosonic supersymmetry ghosts. We propose to look for a worldsheet counterpart of the twist by expanding this action about a constant background for a suitable combination of these ghosts. If the string-level
conjecture holds, such a twist of the complete gauged worldsheet theory
should lead directly to the topological string, without passing through
an auxiliary twisted matter CFT.

In this paper we test this idea for spinning particles coupled to worldline supergravity. The $\mathcal N=2$ particle can be viewed as a point-particle analogue of the RNS superstring, with its two supercharges corresponding to complex combinations of the left- and right-moving worldsheet supercharges. In the $\mathcal N=2$ model, local worldline supersymmetry has two bosonic ghosts. Each of the two supercharges
squares to zero, so a constant background for one ghost gives a twist.
In the RNS string the corresponding supercharge combines the left- and
right-moving supercharges, and it squares to the level-matching
constraint, which vanishes in the zero-mode sector. On the worldline we keep the matter fields, Lagrange multipliers and ghosts explicit so that we can follow the effect of the ghost shift through quantization. Worldline and worldsheet supersymmetry are different
symmetries from local spacetime supersymmetry, and no identification of
their ghosts is assumed. The particle models also allow us to determine how the choice of state space affects the cohomology of the twisted BRST charge.

The $\mathcal N=2$ particle has a Hamiltonian constraint and two odd
constraints whose mixed bracket gives the
Hamiltonian~\cite{Brink:1976sz,Brink:1976uf,Howe:1988ft}. Starting from the $\mathcal N=2$ particle, one can obtain a model with four supercharges in two distinct ways. In the open-like model, introducing a
compatible complex structure splits the $\mathcal N=2$ charges and
allows the four components to be gauged independently, as in K\"ahler
spinning particles~\cite{Marcus:1994mm,Bastianelli:2009eh}. This gives
the $\mathcal N=4_{\mathrm c}$ particle.\footnote{We write
$\mathcal N=4_{\mathrm c}$ and $\mathcal N=4_{\mathrm d}$ for the two
four-charge models of this paper. The subscript c stands for
\emph{complex}, since the model uses a complex structure on the target to
split the $\mathcal N=2$ supercharges. The subscript d stands for
\emph{doubled}, since the model has two flavors of the $\mathcal N=2$
fermions on a real target.} In the closed-like model, we introduce a
flavor index $m=1,2$ on the fermions of the real $\mathcal N=2$
particle. This gives the
$\mathcal N=4_{\mathrm d}$ particle on a real target, without a complex
structure. A compatible complex structure on the target splits each of the four supercharges into holomorphic and antiholomorphic components. Gauging all eight components independently gives the $\mathcal N=8$ particle.

The open-like hierarchy is therefore
$\mathcal N=2\to\mathcal N=4_{\mathrm c}$, while the closed-like
hierarchy is $\mathcal N=4_{\mathrm d}\to\mathcal N=8$. The $\mathcal N=4_{\mathrm c}$ particle has one complex fermion per target direction, while the $\mathcal N=4_{\mathrm d}$ particle has two. A change of basis therefore cannot relate their realizations of the four-charge algebra. The terms \emph{open-like} and \emph{closed-like} refer to the target-space field complexes obtained from the twisted models with the chosen state spaces and R-ghost prescriptions. In the closed-like model, the flavor index labels the two families of worldline fermions. The two flavors are not the left- and
right-moving sectors of the string, and the particle does not capture string oscillators, winding modes or the sum over spin
structures.

Two choices in the quantization of the twisted models determine the
resulting cohomology. First, we use the polynomial representation of the superghost Weyl algebra~\cite{Boffo:2024,Boffo:2025pcm}, in which states are polynomial in the unshifted ghost $\gamma^\dagger$ and in $\beta$, the antighost conjugate to the shifted ghost $\gamma$. Here $\beta$ acts by multiplication and $\gamma$ acts as $\partial_\beta$, so the shift $\gamma\mapsto\gamma+u$ is formally implemented by conjugation with $e^{u\beta}$. That exponential does not
preserve polynomials, so this formal relation does not imply that the
twisted and untwisted complexes have the same cohomology. Second,
we gauge selected R rotations, using BRST methods familiar from
spinning-particle constructions~\cite{Bonezzi:2018,Bonezzi:2020}.
After the shift, the generator $R$ of a gauged R rotation acquires a
term proportional to $u$, so the corresponding constraint takes the form
\[
 T=R+u\beta,
 \qquad u\ne0.
\]
On polynomial states, $T$ is injective, and each class in the quotient by its image has a unique $\beta$-independent representative. The cohomology of the absolute R-ghost complex is therefore the cohomology of this quotient with its induced differential, with the surviving R ghost raising the degree by one. The unshifted R constraints
are instead imposed relatively. These prescriptions are choices of the
quantum state space. The BV action determines the operator algebra and
the BRST charge, but not their representation on states or the
treatment of the R ghosts.

The resulting coefficient complexes are the de~Rham complex
$(\Omega^\bullet(M),d_{\mathrm{dR}})$ for the $\mathcal N=2$ particle,
the Dolbeault complex $(\Omega^{0,\bullet}(M),\bar\partial)$ for the
$\mathcal N=4_{\mathrm c}$ particle, and the polynomial linear BCOV
complex
\[
 \bigl(\operatorname{PV}(M)[t],\ \bar\partial+t\partial_\Omega\bigr)
\]
for the $\mathcal N=8$ particle. Here $\operatorname{PV}(M)$ denotes the space of polyvector-valued $(0,\bullet)$-forms, and $\partial_\Omega$ is the divergence defined by the holomorphic volume form $\Omega$. The degree-two variable $t$ is the surviving superghost combination (Section~\ref{subsec:n8-bcov-interpretation}). In the full worldline BRST complexes, each cohomology class admits a representative with an overall R-ghost factor of $\zeta$ in each open-like model and of $\zeta_1\zeta_2$ in the $\mathcal N=8$ model.

These results use the full shifted BRST charge together with the stated
R prescriptions. In the $\mathcal N=4_{\mathrm c}$ model, shifting the ghost of $\bar q$ leaves a holomorphic supersymmetry sector in the complex before R gauging. The reduction to $\Omega^{0,\bullet}(M)$ also requires imposing the holomorphic R constraint relatively. In the $\mathcal N=8$ model, keeping only the part of the charge linear in the ghost background would give $\bar\partial$ acting on the coefficient of each power of $t$. The term $t\partial_\Omega$ in the BCOV differential comes from the background-independent part of the BRST charge.

The paper is organized as follows. Section~\ref{sec:open-like} develops the cohomology reduction for the $\mathcal N=2$ particle and applies it to the $\mathcal N=4_{\mathrm c}$ model. Section~\ref{sec:closed-like} starts from the $\mathcal N=4_{\mathrm d}$ particle and derives the linear BCOV differential for its $\mathcal N=8$ extension. In both sections, the resulting complexes describe linearized fields and gauge transformations around fixed backgrounds with flat target metrics. Section~\ref{sec:deformed-algebra} extends the construction to curved targets. The open-sector charge algebra closes on any K\"ahler target. In the closed sector, the twisted charge is nilpotent on the relative state space exactly when the K\"ahler metric is Ricci-flat.

\section{Open-like models}
\label{sec:open-like}
We begin with the $\mathcal N=2$ spinning particle, the simplest worldline model that admits a topological twist. This case allows the BV construction and the cohomology reduction to be developed explicitly, providing the basis for the models with more supercharges.

In Section~\ref{subsec:n2-model}, we first construct the twisted BRST charge in the quantum operator algebra, where its nilpotency can be checked directly. We then derive the same charge from a constant shift of a bosonic supersymmetry ghost in the classical BV action, followed by gauge fixing and canonical quantization. This realizes a worldline analogue of the Costello--Li twist~\cite{Costello:2016} and relates the operator construction to a background of the classical BV theory.

Section~\ref{subsec:n2-polynomial-reduction} computes the cohomology of the full twisted charge on polynomial ghost states. Comparing the models with and without R gauging makes the role of the R constraint explicit. For a nonzero ghost background, the ungauged cohomology contains additional copies of de~Rham cohomology labelled by powers of an antighost. Gauging the R symmetry and treating its ghost absolutely removes these copies, leaving the de~Rham complex with a degree shift from the surviving R ghost.

Sections~\ref{subsec:n4-model} and~\ref{subsec:n4-dolbeault-reduction} extend this construction by introducing a compatible complex structure. The supercharges split into holomorphic and antiholomorphic components, whose independent gauging gives the $\mathcal N=4_{\mathrm c}$ particle. An antiholomorphic supersymmetry-ghost background then leads to the Dolbeault complex, with the holomorphic R constraint imposed relatively and the antiholomorphic R ghost treated absolutely. On a Calabi--Yau threefold, the Lie-algebra-valued Dolbeault complex with its BV grading describes linearized holomorphic Chern--Simons theory, the target-space field theory of the open B-model. This relation motivates the term \emph{open-like}.

\subsection{The \texorpdfstring{$\mathcal N=2$}{N=2} particle and its BV twist}
\label{subsec:n2-model}
The $\mathcal N=2$ spinning particle
\cite{Brink:1976sz,Brink:1976uf,Howe:1988ft} describes maps from a
worldline, parametrized by $\tau$, into a target manifold $M$, together with a pair of fermions valued in
the pulled-back tangent bundle, and has local worldline supersymmetry.
We work
in the first-order formalism and take the target to be flat for the
remainder of this section; the coupling to curved target-space geometry
is discussed in Section~\ref{sec:deformed-algebra}. The matter fields
are the coordinates $X^\mu$, $\mu=1,\dots,\dim M$, their conjugate
momenta $P_\mu$, and a pair of complex fermions $\psi^\mu,\psi^\dagger_\mu$.
They are coupled to worldline supergravity, whose multiplet consists of
the einbein $e$ and the gravitini $\chi,\chi^\dagger$; in the
first-order formalism these fields act as Lagrange multipliers
enforcing constraints. The action is given by
\begin{equation}
\label{eq:n2-action}
 S_{\mathcal N=2}
 =\int d\tau\left(
 P_\mu\dot X^\mu
 +\tfrac{i}{2}\psi^\dagger_\mu\dot\psi^\mu
 +\tfrac{i}{2}\psi^\mu\dot\psi^\dagger_\mu
 -\tfrac{e}{2}P^2
 -\tfrac{i}{2}\chi\,\eta^{\mu\nu}\psi^\dagger_\mu P_\nu
 -\tfrac{i}{2}\chi^\dagger\,\psi^\mu P_\mu
 \right).
\end{equation}
The pairs $(X^\mu,P_\nu)$ and $(\psi^\mu,\psi^\dagger_\nu)$ are
canonically conjugate; the kinetic terms fix the Poisson brackets
$\{X^\mu,P_\nu\}_{\mathrm{cl}}=\delta^\mu{}_\nu$ and
$\{\psi^\mu,\psi^\dagger_\nu\}_{\mathrm{cl}}=-i\delta^\mu{}_\nu$.

Varying the action with respect to $e,\chi,\chi^\dagger$ gives the
first-class constraints
\begin{equation}
\label{eq:n2-constraints}
 H:=P^2,
 \qquad
 q:=\psi^\mu P_\mu,
 \qquad
 q^\dagger:=\eta^{\mu\nu}\psi^\dagger_\mu P_\nu .
\end{equation}
We keep $X^\mu,\psi^\mu$ with upper and $P_\mu,\psi^\dagger_\mu$ with
lower indices, so canonical pairs contract without the metric; other
contractions use the flat metric $\eta^{\mu\nu}$. Through the Poisson
bracket, $H$ generates translations in worldline time and
$q,q^\dagger$ generate global worldline supersymmetry transformations.
The multipliers $e,\chi,\chi^\dagger$ gauge these symmetries, making
them local. The Hamiltonian $H$ is
bosonic, while the supercharges $q$ and $q^\dagger$ are odd. They
satisfy the $\mathcal N=2$ worldline superalgebra
\begin{equation}
 \{q,q^\dagger\}_{\mathrm{cl}}=-iH,
 \qquad
 \{q,q\}_{\mathrm{cl}}
 =\{q^\dagger,q^\dagger\}_{\mathrm{cl}}=0,
 \qquad
 \{H,q\}_{\mathrm{cl}}=\{H,q^\dagger\}_{\mathrm{cl}}=0.
\end{equation}
Each supercharge squares to zero on its own, and only the mixed
bracket produces $H$. This allows us to twist the model by giving only
one of the two supersymmetry ghosts a constant background
(Sections~\ref{subsec:n2-operator-deformation}
and~\ref{subsec:n2-background}).

To quantize the model in the BRST formalism we introduce a ghost for
each constraint: the fermionic ghost $c$ for the Hamiltonian
constraint $H$, and the bosonic ghosts $\gamma$ and $\gamma^\dagger$
for the supercharges $q$ and $q^\dagger$. Their conjugate antighosts
are $b$, $\beta$ and $\beta^\dagger$. In the operator
description, the variables $(X^\mu,\psi^\mu,c,\gamma^\dagger,\beta)$
and their canonical partners
$(P_\mu,\psi^\dagger_\mu,b,\beta^\dagger,\gamma)$ generate the graded
operator algebra
\begin{equation}
\label{eq:n2-quantum-algebra}
 [P_\mu,X^\nu]=-i\delta_\mu{}^\nu,
 \qquad
 \{\psi^\dagger_\mu,\psi^\nu\}=\delta_\mu{}^\nu,
 \qquad
 \{c,b\}=1,
 \qquad
 [\gamma^\dagger,\beta^\dagger]=1,
 \qquad
 [\gamma,\beta]=1.
\end{equation}
For the matter variables we use the quantization rule
$[\widehat F,\widehat G\}_{\mathrm{gr}}=i\widehat{\{F,G\}_{\mathrm{cl}}}$,
with $\hbar=1$.\footnote{Here the hat denotes the operator
corresponding to a classical quantity. In the rest of the paper we
suppress the hats and use the same symbol for a classical quantity and
its operator, since which one is meant is clear from the context.} The ghost brackets are the standard ones of BRST
quantization; Section~\ref{subsec:n2-gauge-fixing} derives them from
the kinetic terms of the gauge-fixed BV action. The resulting nilpotent
first-quantized BRST charge\footnote{Whenever it is clear from the context which model we are discussing, we drop the superscript $\mathcal N=2,4,\dots$ and write simply $Q_{\mathrm{BRST}}$.} is
\begin{equation}
\label{eq:n2-quantum-brst}
 Q^{\mathcal N=2}_{\mathrm{BRST}}
 =-cH+\gamma q+\gamma^\dagger q^\dagger+b\gamma^\dagger\gamma,
 \qquad
 \bigl(Q^{\mathcal N=2}_{\mathrm{BRST}}\bigr)^2=0.
\end{equation}
The first three terms pair each constraint with its ghost. The last
term, cubic in the ghosts, encodes the bracket $\{q,q^\dagger\}=H$.
Two supersymmetry transformations close on a time translation, so the
product $\gamma^\dagger\gamma$ multiplies the antighost $b$ of the
Hamiltonian ghost. The nilpotency of $Q_{\mathrm{BRST}}$ can be
checked directly. The relation
$\{q,q^\dagger\}=H$ gives a contribution $\gamma^\dagger\gamma H$ to the
square of the supersymmetry terms, which is cancelled by the
anticommutator of $-cH$ with $b\gamma^\dagger\gamma$.

The model also has a $U(1)$ R symmetry that rotates the complex
fermions $\psi^\mu$ and $\psi^\dagger_\mu$ by opposite phases. To leave
$Q^{\mathcal N=2}_{\mathrm{BRST}}$ invariant, this symmetry must act on the supersymmetry
ghosts as well. The corresponding generator is
\begin{equation}
\label{eq:n2-r-charge}
 R=\psi^\mu\psi^\dagger_\mu-\gamma^\dagger\beta^\dagger+\beta\gamma.
\end{equation}
The supercharges $q,q^\dagger$ have R charges $+1,-1$, while
$\gamma,\gamma^\dagger$ have R charges $-1,+1$. Thus each
term in $Q_{\mathrm{BRST}}$ is R-neutral, and $[R,Q_{\mathrm{BRST}}]=0$.

In Section~\ref{subsec:n2-operator-deformation} we deform
$Q_{\mathrm{BRST}}$ and gauge $R$ directly within this operator algebra.
Sections~\ref{subsec:n2-r-gauging}--\ref{subsec:n2-gauge-fixing} then
derive the resulting operators from a background of the classical BV
theory, Section~\ref{subsec:n2-state-space} specifies the state space
on which they act, and Section~\ref{subsec:n2-polynomial-reduction}
computes their cohomology.

\subsubsection{Deforming the BRST charge}
\label{subsec:n2-operator-deformation}
We first describe the deformation directly in the operator algebra, starting from the BRST charge
$Q_{\mathrm{BRST}}$ of
Eq.~\eqref{eq:n2-quantum-brst}. The matter supercharge $q$ alone
does not anticommute with $Q_{\mathrm{BRST}}$, since its bracket with
the term $\gamma^\dagger q^\dagger$ does not vanish. The BRST
differential acts on $q$ as $\{Q_{\mathrm{BRST}},q\}=\gamma^\dagger H$,
so $q$ alone cannot serve as a deformation. The ghost term
$\gamma^\dagger b$ provides the required completion, because
$\{Q_{\mathrm{BRST}},\gamma^\dagger b\}=-\gamma^\dagger H$ cancels the
matter contribution. A valid deformation is therefore
\begin{equation}
 D:=q+\gamma^\dagger b,\qquad
 D^2=0,\qquad \{Q_{\mathrm{BRST}},D\}=0.
 \label{eq:n2-operator-twisting-differential}
\end{equation}
Let $u$ be a formal bosonic parameter. A twisted BRST differential is
then given by
\begin{equation}
 Q^{\mathcal N=2}_{\mathrm{tw}}(u):=Q^{\mathcal N=2}_{\mathrm{BRST}}+uD
 =-cH+(\gamma+u)q+\gamma^\dagger q^\dagger
   +b\gamma^\dagger(\gamma+u),\qquad
 Q_{\mathrm{tw}}^2=0.
 \label{eq:n2-operator-deformed-charge}
\end{equation}
Thus the deformation is the translation
$\gamma\mapsto\gamma+u$ in the complete BRST charge,
including its cubic ghost term. Initially $u$ has the same
ghost degree as $\gamma$; the twisted grading defined in Eq.~\eqref{eq:twisted-ghost-grading}
below will make it degree zero. From now on we suppress the dependence on $u$
and write simply $Q_{\mathrm{tw}}$.

The remaining terms $-cH+\gamma^\dagger q^\dagger$ of $Q_{\mathrm{tw}}$
are $D$-exact. Since $cq^\dagger$ is even, the
relevant bracket is an ordinary commutator:
\begin{equation}
 \begin{aligned}
 [D,cq^\dagger]
 &=-c\{q,q^\dagger\}+\gamma^\dagger\{b,c\}q^\dagger
   =-cH+\gamma^\dagger q^\dagger,\\
 Q_{\mathrm{tw}}
 &=(u+\gamma)D+[D,cq^\dagger].
 \end{aligned}
 \label{eq:n2-operator-exact-remainder}
\end{equation}
This identity suggests that the cohomology of $Q_{\mathrm{tw}}$ is
governed by the matter supercharge $q$ alone. To make this precise we
need a representation of the operator algebra on states. In
Section~\ref{subsec:n2-state-space} we represent the matter variables
on differential forms, with $\psi^\mu$ acting as $dX^\mu\wedge$ and
$P_\mu$ as $-i\partial_\mu$, so that $q$ acts as $-i\,d_{\mathrm{dR}}$.
The physical states are the cohomology of $Q_{\mathrm{tw}}$ on this
space. In Section~\ref{subsec:n2-noR-cohomology} we show that the
commutator term $[D,cq^\dagger]$ does not affect this cohomology, using
an invertible transformation of the polynomial states. The $\beta$
tower left by that reduction is then removed by gauging the R
symmetry.

The deformation does not commute with the R charge as it stands.
The R charge contains the term $\beta\gamma$, so the shift
$\gamma\mapsto\gamma+u$ adds a term $u\beta$. The charge $R$ itself is
no longer a symmetry of the twisted differential, since
$[R,Q_{\mathrm{tw}}]=uD\ne0$. In the untwisted spinning particle,
gauging $R$ together with a one-dimensional Chern--Simons term is the
standard way to project onto forms of a fixed
degree~\cite{Howe:1988ft,Bastianelli:2005vk}. In the twisted model it instead removes
redundant copies of the cohomology. We will see that for $u\ne0$ the twisted cohomology without R
gauging contains one copy of the de~Rham cohomology for every power of
the bosonic antighost $\beta$. Gauging the shifted R constraint leaves
a single copy, shifted in degree by the R ghost
(Section~\ref{subsec:n2-polynomial-reduction}). We introduce a fermionic R ghost $\zeta$ and
its antighost $b_R$, with $\{\zeta,b_R\}=1$. The unshifted and shifted
R-gauged BRST charges are
\begin{equation}
 Q^{\mathcal N=2}_R=Q^{\mathcal N=2}_{\mathrm{BRST}}+\zeta R,
 \qquad
 Q^{\mathcal N=2}_{\mathrm{tw},R}=Q^{\mathcal N=2}_{\mathrm{BRST}}+uD+\zeta(R+u\beta).
 \label{eq:n2-quantum-twisted-brst}
\end{equation}
In addition to $D^2=\{Q_{\mathrm{BRST}},D\}=0$ from
Eq.~\eqref{eq:n2-operator-twisting-differential}, the canonical algebra
gives
\begin{equation}
 [R,Q_{\mathrm{BRST}}]=0,\qquad [R,D]=D,\qquad
 [Q_{\mathrm{BRST}},\beta]=D,\qquad [D,\beta]=0.
 \label{eq:n2-quantum-identities}
\end{equation}
These identities give
\[
 [R+u\beta,Q_{\mathrm{tw}}]
 =[R,Q_{\mathrm{BRST}}]+u[R,D]+u[\beta,Q_{\mathrm{BRST}}]+u^2[\beta,D]
 =uD-uD=0 .
\]
Together with $Q_{\mathrm{tw}}^2=0$ from
Eq.~\eqref{eq:n2-operator-deformed-charge} and $\zeta^2=0$, this
implies $Q_{\mathrm{tw},R}^2=0$.

We grade the twisted complex by combining the ordinary ghost number
$\operatorname{gh}$, which is $+1$ for $c,\gamma,\gamma^\dagger,\zeta$,
$-1$ for $b,\beta,\beta^\dagger,b_R$ and $0$ for the matter variables,
with the R charge,
\begin{equation}
 \operatorname{gh}_{\mathrm{tw}}=\operatorname{gh}+R .
 \label{eq:twisted-ghost-grading}
\end{equation}
The ghost $\gamma$ has degree zero in this grading, so its constant
background $u$ also has degree zero, and both twisted BRST operators
have degree one.

In the rest of this subsection we derive $Q_{\mathrm{tw}}$ and
$Q_{\mathrm{tw},R}$ from the classical BV theory, expanded around a
constant background for the supersymmetry ghost $\gamma$, followed by
gauge fixing and canonical quantization. This traces the deformation
back to the local supersymmetry of the particle and fixes the
accompanying terms involving the gauge fields and ghosts, including the
shift $R\mapsto R+u\beta$. The operator construction shows that the
deformation exists, and the BV derivation identifies it as the
Costello--Li twist of the gauged particle.

\subsubsection{R gauging and the BV action}
\label{subsec:n2-r-gauging}
At the classical level, the R symmetry of
Eq.~\eqref{eq:n2-r-charge} rotates the complex fermions. Its matter
part is generated by $j=\psi^\mu\psi^\dagger_\mu$, whose nonzero
brackets with the $\mathcal N=2$ supersymmetry generators are
$\{j,q\}_{\mathrm{cl}}=-iq$ and $\{j,q^\dagger\}_{\mathrm{cl}}=iq^\dagger$.

To gauge the R symmetry we introduce a worldline gauge field $a$,
which transforms as $\delta a=\dot\alpha$ under a local R rotation with
parameter $\alpha(\tau)$ and acts as a Lagrange multiplier for $j$.
The gauged action is
\begin{equation}
 S_{0,R}=S_{\mathcal N=2}-\int d\tau\,aj.
 \label{eq:n2-r-gauged-classical}
\end{equation}
The equation of motion of $a$ imposes the constraint $j=0$, which
restricts the states to a subspace of definite R charge.\footnote{One
could add a one-dimensional Chern--Simons
term~\cite{Bastianelli:2005vk,Bastianelli:2013pta} $\kappa\int d\tau\,a$,
which shifts the constraint to $j=\kappa$. The superghost background
used below requires $\kappa=0$, so we omit it.} The ghost of this
gauge symmetry is the R ghost $\zeta$ of
Section~\ref{subsec:n2-operator-deformation}.

We now construct the BV master action. The ghosts $c,\gamma,\gamma^\dagger$ and the R ghost $\zeta$ all have ghost number one. For each field or ghost $\Phi^I$, we introduce an antifield $\Phi_I^*$ with parity $\epsilon(\Phi_I^*)=\epsilon(\Phi^I)+1$ modulo two and ghost number $\operatorname{gh}(\Phi_I^*)=-1-\operatorname{gh}(\Phi^I)$. These antifields
are distinct from the antighosts $b,\beta,\beta^\dagger$ of the BRST
description, which enter the BV formalism as nonminimal fields in
Section~\ref{subsec:n2-gauge-fixing}. We place antifields on the right and use the BV antibracket~\cite{Henneaux:1989jq}
\begin{equation}
 (F,G)_{\mathrm{BV}}
 =-\sum_I\int d\tau\left(
 \frac{\delta_r F}{\delta\Phi^I}
 \frac{\delta_l G}{\delta\Phi_I^*}
 -
 \frac{\delta_r F}{\delta\Phi_I^*}
 \frac{\delta_l G}{\delta\Phi^I}
 \right),
 \qquad
 sF=(S_{\mathrm{BV}},F)_{\mathrm{BV}}.
 \label{eq:n2-bv-antibracket}
\end{equation}
Here $\delta_l$ and $\delta_r$ denote left and right functional derivatives. The minimal BV action of the model without R gauging is
\begin{equation}
 \begin{aligned}
 S_{\mathrm{BV}}^{\mathcal N=2}
 =S_{\mathcal N=2}+\int d\tau\,\Bigl[&
 (-2\dot c-\gamma\chi-\gamma^\dagger\chi^\dagger)e^*\\
 &+(-2c\,\eta^{\mu\nu}P_\nu+\gamma\psi^\mu
    +\gamma^\dagger\eta^{\mu\nu}\psi^\dagger_\nu)X^*_\mu\\
 &+i\gamma^\dagger\eta^{\mu\nu}P_\nu\psi^*_\mu
 +i\gamma P_\mu\psi^{\dagger*\mu}\\
 &-2i\dot\gamma^\dagger\chi^*-2i\dot{\gamma}\chi^{\dagger*}
 +i\gamma\gamma^\dagger c^*\Bigr].
 \end{aligned}
 \label{eq:n2-noR-bv}
\end{equation}

The coefficient of each antifield gives the BRST variation of the
corresponding field or ghost; in particular, the term
$i\gamma\gamma^\dagger c^*$ encodes the mixed bracket $\{q,q^\dagger\}$.
The terms $-2i\dot\gamma^\dagger\chi^*$ and $-2i\dot\gamma\chi^{\dagger*}$
give the BRST variations $s\chi=-2i\dot\gamma^\dagger$ and
$s\chi^\dagger=-2i\dot\gamma$, which follow from the local supersymmetry
transformations of the gravitini by replacing the parameters with the
corresponding ghosts. After R gauging these derivatives become R-covariant.
When the R symmetry is gauged, the BV action becomes
\begin{equation}
 \begin{aligned}
 S_{\mathrm{BV},R}^{\mathcal N=2}
 &=S_{\mathrm{BV}}^{\mathcal N=2}+ S_{R},\\
 S_R
 &=\int d\tau\,\Bigl[
 -aj+\dot\zeta\,a^*
 -2a\gamma^\dagger\chi^*+2a\gamma\chi^{\dagger*}\\
 &\hspace{17mm}
 +i\zeta\bigl(\psi^\mu\psi^*_\mu
 -\psi^\dagger_\mu\psi^{\dagger*\mu}
 +\chi\chi^*-\chi^\dagger\chi^{\dagger*}
 +\gamma^\dagger\gamma^{\dagger*}-\gamma\gamma^*\bigr)\Bigr].
 \end{aligned}
 \label{eq:n2-r-full-bv}
\end{equation}

The terms $-2a\gamma^\dagger\chi^*$ and $2a\gamma\chi^{\dagger*}$ complete the
gravitino transformations to covariant derivatives,
$s\chi=-2i\nabla_\tau^+\gamma^\dagger+i\zeta\chi$ and
$s\chi^\dagger=-2i\nabla_\tau^-\gamma-i\zeta\chi^\dagger$, where
$\nabla_\tau^\pm=\partial_\tau\mp ia$ acts on fields of R charge $\pm1$.
The terms proportional to $\zeta$ give the R rotations of the fermions,
gravitini and supersymmetry ghosts, while $\dot\zeta\,a^*$ represents
the transformation $sa=\dot\zeta$ of the gauge field under a local R
transformation. The BRST transformations are nilpotent off shell and
leave the classical action invariant,\footnote{Invariance holds up to a
total derivative in the Lagrangian.} so both
actions obey the classical master equation $(S,S)_{\mathrm{BV}}=0$. No
terms quadratic in the antifields are required.

\subsubsection{The superghost background}
\label{subsec:n2-background}
To perform a worldline analogue of the Costello--Li
twist~\cite{Costello:2016}, we expand the BV action around a background
in which one supersymmetry ghost has a nonzero constant value. We give
$\gamma$ the constant value $u\ne0$. Since $\gamma$ multiplies $q$ in
the BRST charge, this background selects the supercharge $q$, which
squares to zero.

The values of the remaining fields are not free, since the background
must solve the equations of motion of the BV action. At vanishing
antifields, these equations split into two sets. Varying with respect
to the antifields gives $s\Phi=0$ for every field and ghost, so the
background must be BRST invariant. Varying with respect to the fields
gives the classical equations of motion. A background with vanishing
antifields is therefore a BRST-invariant classical solution.

In a bosonic background all Grassmann-odd fields vanish, in particular
$\psi^\mu$, $\psi^\dagger_\mu$, $\chi$, $\chi^\dagger$, $c$ and $\zeta$.
The odd BRST variations, such as $sX^\mu$, $se$ and $s\gamma$, then
vanish automatically. With $\gamma=u$, the even variations read off
from Eqs.~\eqref{eq:n2-noR-bv} and~\eqref{eq:n2-r-full-bv} contain the
terms
\[
 sc\ni iu\gamma^\dagger,\qquad
 s\psi^\dagger_\mu\ni iuP_\mu,\qquad
 s\chi^\dagger\ni-2i\dot u+2au .
\]
Since $u\ne0$, the condition $sc=0$ requires $\gamma^\dagger=0$, so the
two supersymmetry ghosts cannot both have a nonzero background. The
condition $s\psi^\dagger_\mu=0$ gives $P_\mu=0$, and since $u$ is
constant, $s\chi^\dagger=0$ requires $a=0$. The equation of motion of
$P_\mu$ then reduces to $\dot X^\mu=e\,\eta^{\mu\nu}P_\nu=0$, so $X^\mu$
is constant, and the
constraints $P^2=0$ and $j=0$, together with the remaining classical
equations, are satisfied. The einbein is not fixed by these equations.
A nonvanishing einbein can be brought to a constant by a
reparametrization of $\tau$, and on an open worldline the constant can
then be set to one by rescaling $\tau$, so we choose $e=1$. Thus the
background is
\[
 \gamma=u,\qquad X^\mu=X_0^\mu,\qquad e=1,
\]
with $X_0^\mu$ constant and all other fields and antifields vanishing.
For the model without R gauging, we simply omit $a$ and $\zeta$. The
constant $X_0^\mu$ corresponds to the choice of origin of the target-space
coordinates. It is a modulus of the background, not a restriction on
the theory. The flat-space action does not depend on it, and the
fluctuations around the background include arbitrary worldlines
$X^\mu(\tau)$. Translations shift the modulus, and Lorentz
transformations about $X_0$ leave the background invariant.

We write the full supersymmetry ghost as $u+\gamma$, using $\gamma$
from now on for its fluctuation. This amounts to replacing $\gamma$ by
$\gamma+u$ throughout the BV action. Since the action is at most linear
in $\gamma$, the shifted action is
\[
 S^{\mathcal N=2}_{\mathrm{tw},R}
 =S_{\mathrm{BV},R}^{\mathcal N=2}
      \big|_{\gamma\mapsto\gamma+u}
 =S_{\mathrm{BV},R}^{\mathcal N=2}
      +uS^{\mathcal N=2}_{\mathrm{top},R},
\]
where the coefficients of $u$ without and with R gauging are
\begin{equation}
 \begin{aligned}
 S^{\mathcal N=2}_{\mathrm{top}}
 &=\int d\tau\,\left(
 -\chi e^*+\psi^\mu X_\mu^*
 +iP_\mu\psi^{\dagger*\mu}
 +i\gamma^\dagger c^*\right),\\
 S^{\mathcal N=2}_{\mathrm{top},R}
 &=S^{\mathcal N=2}_{\mathrm{top}}+\int d\tau\,\left(
 2a\chi^{\dagger*}-i\zeta\gamma^*\right).
 \end{aligned}
 \label{eq:n2-top-functional}
\end{equation}
The last two terms arise from shifting $\gamma$ in the R-gauging terms
of Eq.~\eqref{eq:n2-r-full-bv}. They must be kept, since omitting them would
change the twisted R gauge symmetry. A constant translation of $\gamma$
preserves the BV antibracket, so the shifted action satisfies the
classical master equation for every constant $u$. Expanding this
equation in $u$ gives
\[
 \bigl(S_{\mathrm{BV},R}^{\mathcal N=2},
       S_{\mathrm{top},R}\bigr)_{\mathrm{BV}}=0,
 \qquad
 \bigl(S_{\mathrm{top},R},
       S_{\mathrm{top},R}\bigr)_{\mathrm{BV}}=0.
\]

The background is constant in $\tau$ and, with $a=0$, is defined
globally on the worldline. We next gauge fix $S_{\mathrm{tw},R}$ and determine its
quantum BRST operator.

\subsubsection{Gauge fixing and the twisted BRST charge}
\label{subsec:n2-gauge-fixing}
We gauge fix the shifted BV action by imposing $e=1$ and
$\chi=\chi^\dagger=a=0$. For each gauge condition we introduce a
nonminimal pair consisting of an antighost, the BRST antighost of
Section~\ref{subsec:n2-model} rather than a BV antifield, and an
auxiliary field,
\[
 (b,\lambda_e),\qquad
 (\beta^\dagger,\lambda_\chi),\qquad
 (\beta,\lambda_{\chi^\dagger}),\qquad
 (b_R,\lambda_a),
\]
where the last pair is present only when the R symmetry is gauged. The BRST
differential maps each antighost to its auxiliary field and annihilates
the latter. The antighosts $b,b_R$ are fermionic, while
$\beta,\beta^\dagger$ are bosonic; each auxiliary field has the
opposite parity to its antighost. The antighosts have ghost number $-1$
and the auxiliary fields have ghost number zero. We add the nonminimal
action $S_{\mathrm{nm}}$ to $S_{\mathrm{tw},R}$. It pairs each
antighost with its auxiliary field in a trivial BRST doublet, so it
preserves the classical master equation and does not change the
cohomology. It provides the fields needed to write the gauge-fixing
fermion, which we choose as
\begin{align}
 S_{\mathrm{nm}}
 &=\int d\tau\,\left(
 \lambda_e b^*+\lambda_\chi\beta^{\dagger*}
 +\lambda_{\chi^\dagger}\beta^*+\lambda_a b_R^*
 \right),
 \label{eq:n2-nonminimal-action}\\
 \Psi_{\mathrm{gf}}
 &=\int d\tau\,\left[
 \frac{i}{2}b(e-1)-\frac12\beta^\dagger\chi
 -\frac12\beta\chi^\dagger-i b_Ra
 \right].
 \label{eq:n2-gauge-fermion}
\end{align}
We set $\Phi^*=\delta_l\Psi_{\mathrm{gf}}/\delta\Phi$ and eliminate the
auxiliary fields, whose equations impose the gauge conditions. Up to a
total derivative, the resulting actions are
\begin{equation}
 \begin{aligned}
 S_{\mathrm{gf}}
 &=\int d\tau\,\left(
 P_\mu\dot X^\mu-\frac12P^2
 +i\psi^\dagger_\mu\dot\psi^\mu
 +ib\dot c
 +i\beta^\dagger\dot\gamma^\dagger
 +i\beta\dot\gamma
 \right),\\
 S_{\mathrm{gf},R}
 &=S_{\mathrm{gf}}+i\int d\tau\,b_R\dot\zeta,
 \end{aligned}
 \label{eq:n2-gf-action}
\end{equation}
where $S_{\mathrm{gf}}$ is the gauge-fixed action without R gauging. The constant
ghost shift leaves the gauge-fixed action unchanged, because the kinetic term
contains only $\dot\gamma$, which is unaffected since $\dot u=0$, and
the remaining $u$-dependent terms vanish under the gauge-fixing
prescription. The $u$-dependence survives only in the antifield terms
$uS_{\mathrm{top},R}$ of $S_{\mathrm{tw},R}$. They do not contribute to
$S_{\mathrm{gf}}$ in our gauge, but they determine the BRST
transformations $s\Phi=(S_{\mathrm{tw},R},\Phi)_{\mathrm{BV}}$ and hence
the BRST charge below.

The kinetic terms determine the canonical brackets. Upon canonical
quantization with $\hbar=1$, they give the operator
algebra~\eqref{eq:n2-quantum-algebra} of
Section~\ref{subsec:n2-model}, with $\beta$ and $\beta^\dagger$
conjugate to $\gamma$ and $\gamma^\dagger$, respectively. In the
R-gauged model we also obtain $\{\zeta,b_R\}=1$. All other brackets
between the generators of the operator algebra vanish.

After gauge fixing and eliminating the auxiliary fields, the BRST
transformations act on the remaining matter and ghost variables. In
the quantum theory, the BRST transformations become a graded derivation
of the operator algebra. We realize it as the graded commutator with an
operator $Q$, $sF=i[Q,F\}_{\mathrm{gr}}$, chosen to reproduce the
gauge-fixed BRST transformations on the generators. In this paper, quantization
means that Poisson brackets are replaced by the graded
commutators of Eq.~\eqref{eq:n2-quantum-algebra}, and the BRST
transformations by a commutator with $Q$. Here $Q$ is
obtained from $Q_{\mathrm{BRST}}$ or $Q_R$ by replacing $\gamma$ with
$\gamma+u$. These are exactly the
operators $Q_{\mathrm{tw}}$ and $Q_{\mathrm{tw},R}$ of
Eqs.~\eqref{eq:n2-operator-deformed-charge}
and~\eqref{eq:n2-quantum-twisted-brst}, including the shift
$R\mapsto R+u\beta$ of the R constraint. The BV background thus
reproduces the deformation of
Section~\ref{subsec:n2-operator-deformation} and identifies it as the
Costello--Li twist of the gauged particle.

\subsubsection{The polynomial state space}
\label{subsec:n2-state-space}
To represent the operator algebra~\eqref{eq:n2-quantum-algebra} on
states we must choose a polarization. For each canonical pair, one
member acts by multiplication on wave functions and its partner as a
derivative. In the language of geometric quantization, this is a real
polarization. The multiplication operators form a maximal set of
graded-commuting variables, half of each canonical pair, and states are
functions of these variables only. We take
\begin{equation}
 \text{multiplication: }\ X^\mu,\ \psi^\mu,\ c,\ \gamma^\dagger,\ \beta,
 \qquad
 \text{derivative: }\ P_\mu,\ \psi^\dagger_\mu,\ b,\ \beta^\dagger,\ \gamma,
 \label{eq:n2-polarization}
\end{equation}
which is the split already used in
Section~\ref{subsec:n2-model}, and we allow wave functions that are
polynomial in the bosonic multiplication variables $\gamma^\dagger,\beta$.\footnote{In the language
of~\cite{Boffo:2024,Boffo:2025pcm} this is the picture-zero
representation of the superghost Weyl algebra. We do not use other
pictures in this paper.} For $X$ the choice is the usual position
representation, with wave functions on the target space $M$. For the
odd pairs $(\psi,\psi^\dagger)$ and $(c,b)$ the
state spaces are finite-dimensional in these variables, and the two
choices are related by a Grassmann Fourier transform, so the choice is
a convention. For the bosonic ghosts it is not, because a Fourier transform
would map polynomials in $\beta$ to derivatives of delta functions in
$\gamma$, which lie outside the polynomial class. In particular, the
choice for the pair $(\gamma,\beta)$ determines whether the shift can
change the BRST cohomology. If states were polynomials in $\gamma$, the shift
$\gamma\mapsto\gamma+u$ would act on them as the substitution
$f(\gamma)\mapsto f(\gamma+u)$, an invertible map of polynomials, and
the twisted charge would be conjugate to the untwisted one. We
therefore take the antighost $\beta$ of the shifted ghost to act by
multiplication, so that $\gamma=\partial_\beta$. For the unshifted
pair $(\gamma^\dagger,\beta^\dagger)$ we let the ghost $\gamma^\dagger$
act by multiplication. This is a choice of polarization made in the
quantization, the same one used in~\cite{Boffo:2025pcm}. Before
including the R ghost, the
state space is
\begin{equation}
 \mathcal H_{(0)}^{\mathcal N=2}
 =C^\infty(M)\otimes
 \Lambda^\bullet\!\langle\psi^\mu,c\rangle\otimes
 \mathbb C[\gamma^\dagger,\beta]\,
 \lvert0\rangle_0 .
 \label{eq:picture-zero-space}
\end{equation}
Here $\Lambda^\bullet\langle\dots\rangle$ denotes the exterior algebra
generated by the listed odd variables and $\mathbb C[\dots]$ the
polynomial ring in the listed even variables. The vacuum $\lvert0\rangle_0$ is represented by the constant
polynomial $1$; it is annihilated by $\gamma$, $\beta^\dagger$, $b$
and $\psi^\dagger_\mu$. Each state has finite degree in the bosonic variables
$\gamma^\dagger,\beta$, with no common upper bound on the degree;
infinite series and delta-function states are excluded.

The function space on the target is a separate choice of the state
space. We take smooth functions $C^\infty(M)$, but one may instead
impose compact support and use $C^\infty_c(M)$. Both are representations
of the same operator algebra, so the choice belongs to the specification
of the quantum state space rather than to the BV data, in the same way
as the polarization above. It is independent of the superghost
representation and does not affect the mechanism by which the shifted R
constraint removes the $\beta$ tower. All operators used in the
reductions of this paper are differential operators on $M$ and preserve
compact support, so every result holds for either choice. Smooth
functions give the ordinary de~Rham, Dolbeault and BCOV cohomologies,
and compact support gives their compactly supported versions. The two
differ only for noncompact $M$.

The variables
$X^\mu,\psi^\mu,c,\gamma^\dagger,\beta$ act by multiplication, and their
canonical partners act as
\begin{equation}
 \begin{aligned}
 P_\mu&=-i\frac{\partial}{\partial X^\mu},
 &\psi^\dagger_\mu&=\frac{\partial}{\partial\psi^\mu},
 &b&=\frac{\partial}{\partial c},\\
 \gamma&=\frac{\partial}{\partial\beta},
 &\beta^\dagger&=-\frac{\partial}{\partial\gamma^\dagger}.
 \end{aligned}
 \label{eq:picture-zero-derivatives}
\end{equation}
All derivatives with respect to Grassmann variables are left
derivatives. A matter state is a function $\omega(X,\psi)$ of the
coordinates and of the Grassmann variables $\psi^\mu$. Expanding it in
$\psi$,
\[
 \omega(X,\psi)=\sum_{k}\frac{1}{k!}\,\omega_{\mu_1\dots\mu_k}(X)\,
 \psi^{\mu_1}\cdots\psi^{\mu_k},
\]
identifies it with a complex-valued differential form on $M$ under
$\psi^\mu\leftrightarrow dX^\mu$, so that
$C^\infty(M)\otimes\Lambda^\bullet\langle\psi^\mu\rangle=\Omega^\bullet(M)$.
In this language
\[
 q=-i\psi^\mu\frac{\partial}{\partial X^\mu}
   =-i\,d_{\mathrm{dR}}.
\]

The R charge~\eqref{eq:n2-r-charge} acts on the supersymmetry ghosts
as well as on the fermions. In this representation it becomes
\[
 R=\psi^\mu\frac{\partial}{\partial\psi^\mu}
   +\gamma^\dagger\frac{\partial}{\partial\gamma^\dagger}
   +\beta\frac{\partial}{\partial\beta}
 =N_\psi+N_{\gamma^\dagger}+N_\beta ,
\]
where $N_x$ counts powers of $x$, so $R$ counts the total degree in
$\psi^\mu$, $\gamma^\dagger$ and $\beta$. We use this operator
ordering, with multiplication operators to the left of derivatives,
throughout, with no additional constant shift of $R$, so that the
classical constraint $j=0$ becomes $R=0$.

For the R-gauged model we include the R ghost $\zeta$ and allow states
to depend arbitrarily on it. This is the absolute prescription,
explained below. The state space is
\begin{equation}
 \widehat{\mathcal H}_{(0)}^{\mathcal N=2}
 =C^\infty(M)\otimes
 \Lambda^\bullet\!\langle\psi^\mu,c,\zeta\rangle\otimes
 \mathbb C[\gamma^\dagger,\beta]\,
 \lvert0\rangle_0 .
 \label{eq:n2-r-picture-zero-space}
\end{equation}

The ghost $\zeta$ acts by multiplication and its antighost acts as the
formal derivative $b_R=\partial_\zeta$. Moving $\zeta$ to the left in
each monomial, every state can be written uniquely as
\[
 \Phi=f+\zeta g\in\widehat{\mathcal H}_{(0)}^{\mathcal N=2},
 \qquad f,g\in\mathcal H_{(0)}^{\mathcal N=2},
 \qquad b_R\Phi=g .
\]
For a gauged R symmetry there are two standard ways to treat its
ghosts~\cite{Craps:2005wk}. We describe them for the $U(1)$ case relevant here. Before
twisting, the \emph{relative} prescription restricts the
state space to states with $b_R\Phi=0$ and $R\Phi=0$, that is, to
$\zeta$-independent states of vanishing R charge, so that the
constraint is imposed directly on the states. In the \emph{absolute} prescription one keeps arbitrary
$f$ and $g$, and the constraint enters only through the term $\zeta R$
of the differential. The two prescriptions define different complexes.
Here we use the absolute one and compute the cohomology on the entire
state space $\widehat{\mathcal H}_{(0)}^{\mathcal N=2}$, in which the shifted R
constraint enters the differential through the term $\zeta(R+u\beta)$
of Eq.~\eqref{eq:n2-quantum-twisted-brst}. The relative prescription
would require $(R+u\beta)\Phi=0$, which has no nonzero polynomial
solutions (Section~\ref{subsec:n2-full-brst}).

In the twisted grading
$\operatorname{gh}_{\mathrm{tw}}=\operatorname{gh}+R$ of
Section~\ref{subsec:n2-operator-deformation}, the degrees of the state
variables are
\begin{equation}
 \begin{array}{c|ccccc}
 \text{variable}
 &\psi^\mu&c&\gamma^\dagger&\beta&\zeta\\ \hline
 \operatorname{gh}_{\mathrm{tw}}
 &1&1&2&0&1
 \end{array}
 \label{eq:n2-state-degrees}
\end{equation}
Section~\ref{subsec:n2-polynomial-reduction} computes the cohomology of
$Q_{\mathrm{tw}}$ on $\mathcal H_{(0)}^{\mathcal N=2}$ and of
$Q_{\mathrm{tw},R}$ on $\widehat{\mathcal H}_{(0)}^{\mathcal N=2}$, at
fixed $u\ne0$.

\subsection{Reduction to the de~Rham complex}
\label{subsec:n2-polynomial-reduction}

We now compute the cohomology of $Q_{\mathrm{tw}}$ on
$\mathcal H_{(0)}^{\mathcal N=2}$, Eq.~\eqref{eq:picture-zero-space},
and of $Q_{\mathrm{tw},R}$ on $\widehat{\mathcal H}_{(0)}^{\mathcal N=2}$,
Eq.~\eqref{eq:n2-r-picture-zero-space}, at fixed $u\ne0$. Since
$\gamma=\partial_\beta$, one has $e^{-u\beta}\gamma\,e^{u\beta}=\gamma+u$,
so the shift $\gamma\mapsto\gamma+u$ can formally be represented by
conjugation with $e^{u\beta}$. However, this exponential does not
preserve the polynomial state space, so this formal relation does not
imply that the twisted and untwisted operators have the same
cohomology. We return to this point at the end of the section.

The computation has two steps. Without R gauging, an invertible
transformation of the states removes the Hamiltonian and $q^\dagger$
terms from $Q_{\mathrm{tw}}$. The resulting cohomology contains one
copy of the de~Rham cohomology of $M$ for every power of $\beta$.
Gauging the R symmetry adds the term $\zeta(R+u\beta)$ to the
differential. This removes the extra $\beta$ dependence and leaves a
single copy of the de~Rham cohomology, shifted by one in degree by the
R ghost $\zeta$.

\subsubsection{Cohomology of \texorpdfstring{$Q_{\mathrm{tw}}$ on $\mathcal H_{(0)}^{\mathcal N=2}$}{Q\_tw without R gauging}}
\label{subsec:n2-noR-cohomology}
With the identification of matter states with differential forms, the state space
of Eq.~\eqref{eq:picture-zero-space} is
\begin{equation}
 V:=\mathcal H_{(0)}^{\mathcal N=2}
 =\Omega^\bullet(M)\otimes\Lambda\langle c\rangle
   \otimes\mathbb C[\gamma^\dagger,\beta]\,\lvert0\rangle_0,
 \label{eq:n2-noR-state-space}
\end{equation}
on which $q$ acts as $-i\,d_{\mathrm{dR}}$. To compute the cohomology,
we determine the kernel of $Q_{\mathrm{tw}}$ and identify states that
differ by an element of its image.
We first simplify the full BRST operator. In the polynomial
representation, $\gamma=\partial_\beta$ and $b=\partial_c$, and the twisted charge $Q_{\mathrm{tw}}$ given in
Eq.~\eqref{eq:n2-operator-exact-remainder} becomes
\begin{equation}
 Q_{\mathrm{tw}}=A_uD-cH+\gamma^\dagger q^\dagger,
 \qquad
 A_u:=u+\partial_\beta,
 \qquad
 D=q+\gamma^\dagger\partial_c .
 \label{eq:n2-noR-factorization}
\end{equation}
For $u\ne0$ the operator $A_u=u\,(1+\partial_\beta/u)$ is invertible on finite polynomials, because its inverse is the geometric series
\begin{equation}
 A_u^{-1}=\frac1u\sum_{j\ge0}\left(-\frac{\partial_\beta}{u}\right)^j.
 \label{eq:n2-Au-inverse}
\end{equation}
The series terminates on every state because each $\partial_\beta$ lowers the $\beta$ degree.

The remaining terms $-cH+\gamma^\dagger q^\dagger$ equal $[D,cq^\dagger]$. To write them as a
commutator with $A_uD$, we use the fact that $A_u$ commutes with both
$D$ and $cq^\dagger$ and set $L=cq^\dagger A_u^{-1}$. The relations
$\{q,q^\dagger\}=H$, $(q^\dagger)^2=0$ and $\{b,c\}=1$ then give
\begin{equation}
 L^2=0,
 \qquad
 [A_uD,L]=-cH+\gamma^\dagger q^\dagger,
 \qquad
 L(A_uD)L=0,
 \label{eq:n2-noR-similarity-identities}
\end{equation}
where $L^2=0$ because $c^2=0$. Since $L$ is even, expanding the product
gives $(1-L)A_uD(1+L)=A_uD+[A_uD,L]-L(A_uD)L$. The second identity in
Eq.~\eqref{eq:n2-noR-similarity-identities} turns the commutator into
the remaining terms of $Q_{\mathrm{tw}}$, and the third one removes the
cubic term, so
\begin{equation}
 Q_{\mathrm{tw}}
 =A_uD+[A_uD,L]-L(A_uD)L
 =(1-L)A_uD(1+L).
 \label{eq:n2-noR-similarity}
\end{equation}
Thus the invertible transformation $\Phi\mapsto(1+L)\Phi$ carries closed states for $Q_{\mathrm{tw}}$ to closed states for $A_uD$, and carries exact states to exact states. Its inverse is $1-L$, since $L^2=0$. Both transformations preserve the twisted degree.

It remains to compute the cohomology of $A_uD$. Since $A_u$ commutes
with $D$ and is invertible on finite polynomials, $A_uD$ and $D$ have
the same kernel and the same image. The kernels match because $A_u$ is
injective, $\ker(A_uD)=\ker D$, and the images match because $A_u$ is
surjective,
$\operatorname{im}(A_uD)=\operatorname{im}(DA_u)=\operatorname{im}D$. It is therefore enough to compute the
cohomology of $D=q+\gamma^\dagger\partial_c$. The ghost part
$\gamma^\dagger\partial_c$ pairs each monomial containing $c$ with one
containing an additional power of $\gamma^\dagger$,
\[
 \gamma^\dagger\partial_c\bigl(c(\gamma^\dagger)^m\beta^n\bigr)
 =(\gamma^\dagger)^{m+1}\beta^n,
 \qquad
 \gamma^\dagger\partial_c\bigl((\gamma^\dagger)^m\beta^n\bigr)=0,
\]
so a polynomial is closed under $\gamma^\dagger\partial_c$ only if it is
independent of $c$, and every closed polynomial with a nonvanishing
power of $\gamma^\dagger$ is exact. The
cohomology of the ghost part on $\Lambda\langle c\rangle\otimes
\mathbb C[\gamma^\dagger,\beta]$ is therefore $\mathbb C[\beta]$,
represented by polynomials in $\beta$ alone. The matter part
$q=-i\,d_{\mathrm{dR}}$ acts on the separate tensor factor
$\Omega^\bullet(M)$. Since $\beta$ has twisted degree zero, we obtain
\begin{equation}
 H^k\bigl(V,Q_{\mathrm{tw}}\bigr)
 \cong H^k_{\mathrm{dR}}(M)\otimes\mathbb C[\beta],
 \qquad\text{for }u\ne0.
 \label{eq:n2-noR-full-tower}
\end{equation}
For a closed form $\omega\in\Omega^k(M)$, the corresponding classes are
represented by the states $(1-L)\beta^n\omega$, $n\ge0$. The tensor
product in Eq.~\eqref{eq:n2-noR-full-tower} describes only the
graded vector space of classes, since multiplication by $\beta$ does not
commute with the full charge, as $[Q_{\mathrm{tw}},\beta]=D$.
Without R gauging, the twisted model therefore contains one copy of
the de~Rham cohomology for every power of $\beta$.

\subsubsection{Cohomology of \texorpdfstring{$Q_{\mathrm{tw},R}$ on $\widehat{\mathcal H}_{(0)}^{\mathcal N=2}$}{Q\_tw,R with R gauging}}
\label{subsec:n2-full-brst}
We now show how gauging the R symmetry removes the extra copies of de~Rham cohomology found in Section~\ref{subsec:n2-noR-cohomology}. We keep $u\ne0$ fixed and use the polynomial state space $V$ of Eq.~\eqref{eq:n2-noR-state-space}. Every state in the absolute complex has the form $\Phi=f+\zeta g\in\widehat{\mathcal H}_{(0)}^{\mathcal N=2}$, with $f,g\in V$. The twisted charge $Q_{\mathrm{tw},R}$ of Eq.~\eqref{eq:n2-quantum-twisted-brst} acts on the states as
\begin{equation}
 Q_{\mathrm{tw},R}(f+\zeta g)
 =Q_{\mathrm{tw}}f+\zeta\bigl(Tf-Q_{\mathrm{tw}}g\bigr),
 \qquad T=R+u\beta.
 \label{eq:n2-shifted-constraint}
\end{equation}
The minus sign comes from moving the odd operator $Q_{\mathrm{tw}}$ past the odd ghost $\zeta$. Thus $\Phi$ is closed exactly when
\begin{equation}
 Q_{\mathrm{tw}}f=0,
 \qquad Tf=Q_{\mathrm{tw}}g.
 \label{eq:n2-gauged-closedness}
\end{equation}
We show that every class has a representative $\zeta g_0$ with $g_0$ independent of $\beta$ and $Q_{\mathrm{tw}}$-closed, unique up to $Q_{\mathrm{tw}}$-exact terms, and then compute the cohomology of $Q_{\mathrm{tw}}$ on such states, which gives a single copy of de~Rham cohomology shifted in degree by $\zeta$.

We begin with the action of $T$. In the chosen representation,
\[
 T=R+u\beta=N_\psi+N_{\gamma^\dagger}+N_\beta+u\beta.
\]
The number operators preserve the power of $\beta$, while the last term raises it by one. If $h\in V$ is nonzero, of degree $n$ in $\beta$, with leading coefficient $h_n$, then $Th$ has degree $n+1$ in $\beta$, with leading coefficient $uh_n$. Since $u\ne0$, this coefficient cannot vanish. Thus $T$ is injective. Moreover, every nonzero element of its image $TV=\{Th:h\in V\}$ has positive degree in $\beta$ and therefore cannot be independent of $\beta$.

Suppose that $g\in V$ has degree $n\geq1$ in $\beta$, with leading coefficient $g_n$. Applying the $u\beta$ term of $T$ to $\beta^{n-1}g_n/u$ results in $\beta^n g_n$, while the number operators contribute only terms of degree at most $n-1$. Thus
\[
 g_1:=g-T\left(\frac{\beta^{n-1}g_n}{u}\right)\in V
\]
has degree at most $n-1$ in $\beta$.

We now iterate. Applying the same step to $g_1$ gives $g_2$ of degree at most $n-2$, and so on. After at most $n$ steps we reach a $\beta$-independent polynomial $g_0$. Let $h\in V$ be the sum of the polynomials to which $T$ was applied. By linearity of $T$, the total subtraction is $Th$, so $g=Th+g_0$. If $g$ is already independent of $\beta$, we simply take $h=0$ and $g_0=g$.

We will use this decomposition to compare the $TV$ and $\beta$-independent components of the second closedness condition in Eq.~\eqref{eq:n2-gauged-closedness}. For this comparison, the decomposition must be unique. Suppose that $g=Th+g_0=Th'+g_0'$, where $g_0$ and $g_0'$ are independent of $\beta$. Then $T(h-h')=g_0'-g_0$. The left-hand side belongs to $TV$, while the right-hand side is independent of $\beta$. The degree argument above shows that no nonzero element of $TV$ can be independent of $\beta$, so both sides must vanish. Thus $g_0=g_0'$, and injectivity of $T$ gives $h=h'$. We have therefore proved
\begin{equation}
 V=TV\oplus V_0,
 \qquad V_0=\ker\partial_\beta.
 \label{eq:n2-polynomial-division}
\end{equation}
Every state $g\in V$ therefore splits uniquely as $g=Th+g_0$, with
$h\in V$ and $g_0\in V_0$.

Terms of positive degree in $\beta$ can contribute to the $V_0$
component $g_0$ of $g$, its projection onto $V_0$ along $TV$. For
example, if $v\in V_0$, then $\beta v=u^{-1}Tv-u^{-1}Rv$, so the $V_0$
component of $\beta v$ is $-u^{-1}Rv$.

Both subspaces $TV$ and $V_0$ are preserved by $Q_{\mathrm{tw}}$. The commutation relation $[Q_{\mathrm{tw}},T]=0$ gives $Q_{\mathrm{tw}}(Th)=T(Q_{\mathrm{tw}}h)$, so $TV$ is preserved. Since $Q_{\mathrm{tw}}$ contains $\partial_\beta$ but no multiplication by $\beta$, it also preserves $V_0$.

Now let $\Phi=f+\zeta g$ be $Q_{\mathrm{tw},R}$-closed and write $g=Th+g_0$. Subtracting the exact state $Q_{\mathrm{tw},R}h$ from $\Phi$ removes $\zeta Th$ and leaves $(f-Q_{\mathrm{tw}}h)+\zeta g_0$. The second closedness condition in Eq.~\eqref{eq:n2-gauged-closedness} gives
\[
 Tf=Q_{\mathrm{tw}}(Th+g_0)
   =TQ_{\mathrm{tw}}h+Q_{\mathrm{tw}}g_0,
\]
or equivalently,
\[
 T\bigl(f-Q_{\mathrm{tw}}h\bigr)=Q_{\mathrm{tw}}g_0.
\]
The left-hand side belongs to $TV$ and the right-hand side belongs to $V_0$. Their intersection is zero, so $Q_{\mathrm{tw}}g_0=0$. Injectivity of $T$ then gives $f=Q_{\mathrm{tw}}h$. Substituting these results into $\Phi$, we obtain
\begin{equation}
 \Phi=Q_{\mathrm{tw}}h+\zeta(Th+g_0)
      =Q_{\mathrm{tw},R}h+\zeta g_0,
 \qquad Q_{\mathrm{tw}}g_0=0.
 \label{eq:n2-gauged-representative-reduction}
\end{equation}
By Eq.~\eqref{eq:n2-gauged-representative-reduction}, $\Phi$ and $\zeta g_0$ differ by the BRST-exact state $Q_{\mathrm{tw},R}h$ and therefore represent the same cohomology class, so every class has a representative $\zeta g_0$ with $g_0\in V_0$ closed under $Q_{\mathrm{tw}}$.
We must also check which of these representatives are exact. Suppose
\[
 \zeta g_0=Q_{\mathrm{tw},R}(F+\zeta G),
 \qquad F,G\in V.
\]
Comparing the coefficients of $\zeta$ gives $g_0=TF-Q_{\mathrm{tw}}G$. To isolate the component of this equation in $V_0$, write $G=T\ell+G_0$, where $\ell\in V$ and $G_0\in V_0$. Using $[Q_{\mathrm{tw}},T]=0$, we obtain
\[
 g_0=T\bigl(F-Q_{\mathrm{tw}}\ell\bigr)-Q_{\mathrm{tw}}G_0.
\]
The first term on the right belongs to $TV$, whereas $g_0$ and $Q_{\mathrm{tw}}G_0$ belong to $V_0$. Uniqueness of the decomposition therefore gives $g_0=-Q_{\mathrm{tw}}G_0$. Conversely, this equation implies
\[
 \zeta g_0=Q_{\mathrm{tw},R}(\zeta G_0).
\]
Thus $\zeta g_0$ is exact in the full R-gauged complex if and only if $g_0$ is exact in $(V_0,Q_{\mathrm{tw}})$. Applying this to the difference of two representatives shows that they define the same full BRST class exactly when their $g_0$ parts differ by a $Q_{\mathrm{tw}}$-exact state in $V_0$. Since $\zeta$ has twisted degree one, we obtain
\begin{equation}
 H^k\bigl(\widehat{\mathcal H}_{(0)}^{\mathcal N=2},Q_{\mathrm{tw},R}\bigr)
 \cong H^{k-1}\bigl(V_0,Q_{\mathrm{tw}}\bigr),
 \qquad\text{for }u\ne0.
 \label{eq:n2-full-brst-beta-reduction}
\end{equation}
It remains to compute the cohomology on $V_0$. Since states in $V_0$ are independent of $\beta$, $\partial_\beta$ annihilates them and $A_u$ acts as multiplication by $u$. The similarity transformation of Eq.~\eqref{eq:n2-noR-similarity} therefore restricts to
\[
 Q_{\mathrm{tw}}\big|_{V_0}=(1-L_0)\,uD\,(1+L_0),
 \qquad L_0=\frac{cq^\dagger}{u},
 \qquad D=q+\gamma^\dagger\partial_c.
\]
Here $L_0$ and $D$ map $V_0$ to itself. The maps $1\pm L_0$ are inverse to each other, and multiplication of a differential by the nonzero scalar $u$ does not change its cohomology. We can therefore use the same ghost calculation as before. The operator $\gamma^\dagger\partial_c$ leaves only the constant class in $\Lambda\langle c\rangle\otimes\mathbb C[\gamma^\dagger]$, while $q=-i\,d_{\mathrm{dR}}$ acts on the matter forms. There is now no $\beta$ variable to supply extra copies. Hence
\begin{equation}
 H^k\bigl(\widehat{\mathcal H}_{(0)}^{\mathcal N=2},Q_{\mathrm{tw},R}\bigr)
 \cong H_{\mathrm{dR}}^{k-1}(M),
 \qquad\text{for }u\ne0.
 \label{eq:n2-full-gauged-de-rham}
\end{equation}
To obtain a representative in the full R-gauged complex, we apply the inverse transformation $1-L_0$ to a closed form $\omega\in\Omega^{k-1}(M)$ and multiply by $\zeta$. This gives the class
\[
 \biggl[\zeta\left(1-\frac{cq^\dagger}{u}\right)\omega\biggr]
 \in H^k\bigl(\widehat{\mathcal H}_{(0)}^{\mathcal N=2},Q_{\mathrm{tw},R}\bigr).
\]

The reduction to $V_0$ in Eq.~\eqref{eq:n2-full-brst-beta-reduction} applies whenever $T\colon V\to V$ is injective, commutes with the differential, and gives a decomposition $V=TV\oplus V_0$ with $V_0$ preserved by that differential. We will use it again in Sections~\ref{subsec:n4-antiholomorphic-cone} and~\ref{subsec:n8-gauged-reduction}.
\paragraph{Role of the state space.}
\label{par:n2-polynomial-domain}
Both cohomology computations rely on the choice of the polynomial ring in $\beta$, rather than formal power series, in the state space $V$ of Eq.~\eqref{eq:n2-noR-state-space}. The series~\eqref{eq:n2-Au-inverse} terminates on every polynomial state, but not on general power series. Likewise, the successive subtractions used to construct $g=Th+g_0$ terminate because $g$ has finite degree in $\beta$. The formal conjugation identities
\[
 Q_{\mathrm{tw}}=e^{-u\beta}Q_{\mathrm{BRST}}e^{u\beta},
 \qquad
 Q_{\mathrm{tw},R}=e^{-u\beta}Q_Re^{u\beta}
\]
would identify the twisted and untwisted cohomologies on an enlarged state space preserved by the BRST operators and by multiplication by both $e^{u\beta}$ and $e^{-u\beta}$. On the polynomial state space used here, multiplication by $e^{\pm u\beta}$ does not preserve the space. These formal identities therefore do not establish an equivalence between the twisted and untwisted polynomial complexes.
The computation also depends on the absolute R-ghost prescription. In the relative complex, the conditions $b_R\Phi=0$ and $T\Phi=0$ would leave only the zero polynomial state, since $T$ is injective for $u\ne0$. The surviving representatives $\zeta g_0$ belong to the absolute complex, so with the relative prescription the twisted model would have no cohomology at all.

\subsection{The \texorpdfstring{$\mathcal N=4_{\mathrm c}$}{N=4c} spinning particle}
\label{subsec:n4-model}
We now extend the construction to a target with a complex structure.
Splitting the $\mathcal N=2$ supercharges into holomorphic and
antiholomorphic parts gives four charges, and gauging them
independently defines the $\mathcal N=4_{\mathrm c}$ spinning particle.
With the R prescriptions of Section~\ref{subsec:n4-antiholomorphic-cone},
its twisted BRST complex reduces to the Dolbeault complex, as we will
show in Eq.~\eqref{eq:n4-gauged-dolbeault-complex}.

\subsubsection{The \texorpdfstring{$\mathcal N=4_{\mathrm c}$}{N=4c} action and operator algebra}
\label{subsec:n4-classical-model}
We take the target $M$ to be even-dimensional with a constant complex
structure, and choose a flat Hermitian metric, given in complex
coordinates by
\[
 ds^2=2h_{k\bar l}\,dX^k d\bar X^{\bar l},
\]
with $h_{k\bar l}$ constant. The coordinates $X^\mu$, momenta $P_\mu$
and fermions $\psi^\mu,\psi^\dagger_\mu$ split into holomorphic and
antiholomorphic components. We write the antiholomorphic coordinates
and momenta as $\bar X^{\bar k}$ and $\bar P_{\bar k}$, and the
fermions as $\psi^k,\bar\psi^{\bar k}$ with canonical partners
$\psi^\dagger_k,\bar\psi^\dagger_{\bar k}$.
As in Section~\ref{subsec:n2-model}, fields carry upper and their
canonical partners lower indices, and contractions between holomorphic
and antiholomorphic indices use the Hermitian metric $h$. The fermion
brackets inherited from the $\mathcal N=2$ model are
\begin{equation}
 \{\psi^k,\psi^\dagger_l\}_{\mathrm{cl}}=-i\delta^k{}_l,
 \qquad
 \{\bar\psi^{\bar k},\bar\psi^\dagger_{\bar l}\}_{\mathrm{cl}}
 =-i\delta^{\bar k}{}_{\bar l}.
 \label{eq:n4-fermion-brackets}
\end{equation}
The two $\mathcal N=2$ supercharges split into four charges,
\begin{align}
 q&=\psi^kP_k,
 &\bar q&=\bar\psi^{\bar k}\bar P_{\bar k},
 \label{eq:n4-q-split}\\
 q^\dagger&=h^{k\bar l}\psi^\dagger_k\bar P_{\bar l},
 &\bar q^\dagger&=h^{l\bar k}\bar\psi^\dagger_{\bar k}P_l.
 \label{eq:n4-qbar-split}
\end{align}
Each daggered charge is the adjoint of the corresponding undaggered
one, and exchanging holomorphic and antiholomorphic indices maps $q$ to
$\bar q$ and $q^\dagger$ to $\bar q^\dagger$. By
Eq.~\eqref{eq:n4-fermion-brackets}, a charge has a nonvanishing bracket
only with its own adjoint. After quantization $q$ and $\bar q$
represent $-i\partial$ and $-i\bar\partial$, and
$q^\dagger$ and $\bar q^\dagger$ their adjoints. The gravitini, ghosts
and R fields introduced below carry a label
$A\in\{\partial,\bar\partial\}$ indicating whether they belong to the
pair $(q,q^\dagger)$ or to $(\bar q,\bar q^\dagger)$. With
$H=2h^{k\bar l}P_k\bar P_{\bar l}$, the charges satisfy the classical
algebra
\begin{equation}
 \begin{gathered}
 \{q,q^\dagger\}_{\mathrm{cl}}
 =\{\bar q,\bar q^\dagger\}_{\mathrm{cl}}
 =-\frac{i}{2}H,
 \qquad
 \{q,\bar q^\dagger\}_{\mathrm{cl}}
 =\{\bar q,q^\dagger\}_{\mathrm{cl}}=0,
 \end{gathered}
 \label{eq:n4-constraint-algebra}
\end{equation}
while the brackets among $q,\bar q$, among $q^\dagger,\bar q^\dagger$,
and of $H$ with all four charges vanish. In this and the following
sections $q$ denotes the holomorphic charge. The $\mathcal N=2$ charges
of Section~\ref{subsec:n2-model}, written in complex coordinates, are
the sums $q+\bar q$ and $q^\dagger+\bar q^\dagger$, with
$\{q+\bar q,q^\dagger+\bar q^\dagger\}_{\mathrm{cl}}=-iH$ as before.

To define the $\mathcal N=4_{\mathrm c}$ model, we gauge the four split charges
independently, introducing gravitini $\chi_A,\chi_A^\dagger$ for each
$A=\partial,\bar\partial$. The classical action is
\begin{equation}
 \begin{aligned}
 S_0^{\mathcal N=4_{\mathrm c}}=\int d\tau\Big[
 &P_k\dot X^k+\bar P_{\bar k}\dot{\bar X}^{\bar k}
 +i\psi^\dagger_k\dot\psi^k
 +i\bar\psi^\dagger_{\bar k}\dot{\bar\psi}^{\bar k}
 -\frac e2H\\
 &-\frac i2
       \bigl(\chi_\partial^\dagger q+\chi_\partial q^\dagger
       +\chi_{\bar\partial}^\dagger\bar q
       +\chi_{\bar\partial}\bar q^\dagger\bigr)
 \Big].
 \end{aligned}
 \label{eq:n4-classical-action}
\end{equation}
Setting $\chi_\partial=\chi_{\bar\partial}=\chi$ and
$\chi_\partial^\dagger=\chi_{\bar\partial}^\dagger=\chi^\dagger$
recovers the $\mathcal N=2$ action~\eqref{eq:n2-action}, up to a total
derivative in the fermionic kinetic term. The difference between the
two models is which constraints are imposed. In the $\mathcal N=2$
model only the sums $q+\bar q$ and $q^\dagger+\bar q^\dagger$ are
constraints, whereas in
the $\mathcal N=4_{\mathrm c}$ model the gravitini are varied independently, so
each of the four split charges is imposed separately.

\subsubsection{R gauging and the BV action}
\label{subsec:n4-bv-action}
The complex structure on the target extends the $U(1)$ R symmetry of
the $\mathcal N=2$ model to a $U(1)\times U(1)$ R symmetry, since the
holomorphic and antiholomorphic fermions can now be rotated
independently. The two commuting R charges are
\[
 j=\psi^k\psi^\dagger_k,
 \qquad
 \bar j=\bar\psi^{\bar k}\bar\psi^\dagger_{\bar k},
\]
whose sum $j+\bar j$ is the R charge of the $\mathcal N=2$ model.
We gauge both rotations, introducing
worldline gauge fields $a_A$, $A=\partial,\bar\partial$, which act as
Lagrange multipliers for the R charges. The gauged classical action is
\begin{equation}
 S_{0,R}^{\mathcal N=4_{\mathrm c}}
 =S_0^{\mathcal N=4_{\mathrm c}}
 -\int d\tau\,\bigl(a_\partial j+a_{\bar\partial}\bar j\bigr).
 \label{eq:n4-r-gauged-classical}
\end{equation}
We introduce the fermionic ghost $c$, which corresponds to the
Hamiltonian constraint, the bosonic ghosts
$\gamma_\partial,\gamma_\partial^\dagger$ for the symmetries generated
by $q,q^\dagger$ and $\gamma_{\bar\partial},\gamma_{\bar\partial}^\dagger$
for those generated by $\bar q,\bar q^\dagger$, and the fermionic ghosts
$\zeta_A$ for the two R symmetries. We also write
$\eta:=\zeta_\partial$ and $\zeta:=\zeta_{\bar\partial}$. Under the R
rotation labeled by $A$, the ghosts $\gamma_A$ and $\gamma_A^\dagger$
have charges $-1$ and $+1$, and they are neutral under the other R
rotation. Using the conventions of Section~\ref{subsec:n2-r-gauging}
and placing the antifields on the right, the minimal BV action is
$S_{\mathrm{BV},R}^{\mathcal N=4_{\mathrm c}}
=S_{0,R}^{\mathcal N=4_{\mathrm c}}+\int d\tau\sum_\Phi(s\Phi)\,\Phi^*$,
where the sum runs over all fields and ghosts and
\begingroup\setlength{\jot}{1pt}
\begin{equation}
 \begin{aligned}
 se&=-2\dot c-\tfrac12\textstyle\sum_A
       (\gamma_A\chi_A+\gamma_A^\dagger\chi_A^\dagger),
 &sc&=\tfrac i2\textstyle\sum_A\gamma_A\gamma_A^\dagger,\\
 sX^k&=-2c\,h^{k\bar l}\bar P_{\bar l}+\gamma_\partial\psi^k
       +\gamma_{\bar\partial}^\dagger h^{k\bar l}\bar\psi^\dagger_{\bar l},
 &s\bar X^{\bar k}&=-2c\,h^{l\bar k}P_l+\gamma_{\bar\partial}\bar\psi^{\bar k}
       +\gamma_\partial^\dagger h^{l\bar k}\psi^\dagger_l,\\
 s\psi^k&=i\gamma_\partial^\dagger h^{k\bar l}\bar P_{\bar l}
       +i\eta\psi^k,
 &s\bar\psi^{\bar k}&=i\gamma_{\bar\partial}^\dagger h^{l\bar k}P_l
       +i\zeta\bar\psi^{\bar k},\\
 s\psi^\dagger_k&=i\gamma_\partial P_k-i\eta\psi^\dagger_k,
 &s\bar\psi^\dagger_{\bar k}&=i\gamma_{\bar\partial}\bar P_{\bar k}
       -i\zeta\bar\psi^\dagger_{\bar k},\\
 s\chi_A&=-2i\nabla_{\tau A}^+\gamma_A^\dagger+i\zeta_A\chi_A,
 &s\chi_A^\dagger&=-2i\nabla_{\tau A}^-\gamma_A-i\zeta_A\chi_A^\dagger,\\
 s\gamma_A&=-i\zeta_A\gamma_A,\quad
 s\gamma_A^\dagger=i\zeta_A\gamma_A^\dagger,
 &sa_A&=\dot\zeta_A,\quad s\zeta_A=0.
 \end{aligned}
 \label{eq:n4-full-bv-action}
\end{equation}
\endgroup

Here $\nabla_{\tau A}^\pm=\partial_\tau\mp ia_A$, as in the
$\mathcal N=2$ model. The action is linear in the antifields. This is possible
because the Hamiltonian constraint algebra closes off shell with
constant structure coefficients, so the BRST transformations are
nilpotent off shell. Together with the invariance of the classical
action up to a total derivative, this gives the classical master
equation. The BV action without R gauging is obtained by omitting the
terms containing $a_A$ or $\zeta_A$.

\subsubsection{The antiholomorphic ghost background}
\label{subsec:n4-flat-bv-twist}
We now apply the twist of
Section~\ref{subsec:n2-background} to the $\mathcal N=4_{\mathrm c}$ model. Since
the four charges are gauged separately, we can choose which of them to
twist with. We give the antiholomorphic supersymmetry ghost
$\gamma_{\bar\partial}$ the constant background $u\ne0$ and set all
other ghost backgrounds to zero. Other choices are possible. The
condition $sc=0$ only forbids nonzero backgrounds for both $\gamma_A$
and $\gamma_A^\dagger$ with the same $A$. A background for
$\gamma_\partial$ alone gives the holomorphic counterpart of the model
below. Backgrounds for both $\gamma_\partial$ and $\gamma_{\bar\partial}$
select a combination of $q$ and $\bar q$, which reproduces the de~Rham
twist of Section~\ref{subsec:n2-model} when the two backgrounds are
equal. We choose $\gamma_{\bar\partial}$ because this is the choice relevant
for the B-model. The
background $\gamma_{\bar\partial}=u$ selects the charge
$\bar q$, which squares to zero and will become the Dolbeault
operator $\bar\partial$. As in the $\mathcal N=2$ case, the BV equations
of motion are then solved by constant $X$, $e=1$ and vanishing momenta,
fermions, gravitini, R connections and antifields.

Writing the full ghost as $u+\gamma_{\bar\partial}$, the shifted
minimal BV action is
\[
 S_{\mathrm{tw},R}^{\mathcal N=4_{\mathrm c}}
 =S_{\mathrm{BV},R}^{\mathcal N=4_{\mathrm c}}
       \big|_{\gamma_{\bar\partial}
                    \mapsto\gamma_{\bar\partial}+u}
 =S_{\mathrm{BV},R}^{\mathcal N=4_{\mathrm c}}+uS^{\mathcal N=4_{\mathrm c}}_{\mathrm{top},R},
\]
which is exactly linear in $u$ because the action is linear in
$\gamma_{\bar\partial}$. The coefficient of $u$ is
\begin{align}
 S^{\mathcal N=4_{\mathrm c}}_{\mathrm{top}}
 &=\int d\tau\,\left[
 \bar\psi^{\bar k}\bar X^*_{\bar k}
 +i\bar P_{\bar k}\bar\psi^{\dagger *\bar k}
 -\frac12\chi_{\bar\partial}e^*
 +\frac{i}{2}\gamma_{\bar\partial}^\dagger c^*\right],
 \label{eq:n4-twist-bv-generator}\\
 S^{\mathcal N=4_{\mathrm c}}_{\mathrm{top},R}
 &=S^{\mathcal N=4_{\mathrm c}}_{\mathrm{top}}+\int d\tau\,
 \left(2a_{\bar\partial}\chi_{\bar\partial}^{\dagger *}
 -i\zeta\gamma_{\bar\partial}^*\right),
 \label{eq:n4-r-twist-bv-generator}
\end{align}
where $S^{\mathcal N=4_{\mathrm c}}_{\mathrm{top}}$ is the part present without R gauging. Since
the constant translation preserves the BV antibracket, expanding the
master equation of the shifted action in $u$ gives
\[
 (S_{\mathrm{BV},R}^{\mathcal N=4_{\mathrm c}},
       S^{\mathcal N=4_{\mathrm c}}_{\mathrm{top},R})_{\mathrm{BV}}=0,
 \qquad
 (S^{\mathcal N=4_{\mathrm c}}_{\mathrm{top},R},S^{\mathcal N=4_{\mathrm c}}_{\mathrm{top},R})_{\mathrm{BV}}=0.
\]

\subsubsection{Gauge fixing and the twisted BRST charge}
\label{subsec:n4-gauge-fixing}
We gauge fix exactly as in Section~\ref{subsec:n2-gauge-fixing},
with one nonminimal pair for each Lagrange multiplier,
\[
 (b,\lambda_e),\qquad
 (\beta_A^\dagger,\lambda_{\chi_A}),\qquad
 (\beta_A,\lambda_{\chi_A^\dagger}),\qquad
 (b_{R,A},\lambda_{a_A}).
\]
The first entry of each pair is an antighost of ghost number $-1$ and
the second an auxiliary field of ghost number zero; the antighosts
$\beta_A,\beta_A^\dagger$ are bosonic, while $b,b_{R,A}$ are fermionic.
We add the nonminimal action
\[
 S_{\mathrm{nm}}
 =\int d\tau\left[
   \lambda_e b^*
   +\sum_A\left(
      \lambda_{\chi_A}\beta_A^{\dagger *}
      +\lambda_{\chi_A^\dagger}\beta_A^*
      +\lambda_{a_A}b_{R,A}^*
   \right)\right]
\]
to $S_{\mathrm{tw},R}^{\mathcal N=4_{\mathrm c}}$ and choose the
gauge-fixing fermion
\begin{equation}
 \Psi_{\mathrm{gf}}=\int d\tau\,\left[
 \frac{i}{2}b(e-1)
 -\frac12\sum_A
       (\beta_A\chi_A^\dagger+\beta_A^\dagger\chi_A)
 -i\sum_A b_{R,A}a_A\right].
 \label{eq:n4-gauge-fermion}
\end{equation}
Eliminating the auxiliary fields imposes $e=1$ and
$\chi_A=\chi_A^\dagger=a_A=0$. The gauge-fixed action is
\begin{equation}
 \begin{aligned}
 S_{\mathrm{gf},R}^{\mathcal N=4_{\mathrm c}}
 =\int d\tau\,\Big[
 &P_k\dot X^k+\bar P_{\bar k}\dot{\bar X}^{\bar k}
 +i\psi^\dagger_k\dot\psi^k
 +i\bar\psi^\dagger_{\bar k}\dot{\bar\psi}^{\bar k}
 -\frac12H+ib\dot c\\
 &+i\sum_A\left(
       \beta_A\dot\gamma_A
       +\beta_A^\dagger\dot\gamma_A^\dagger
       +b_{R,A}\dot\zeta_A\right)
 \Big],
 \end{aligned}
 \label{eq:n4-gf-action}
\end{equation}
so $\beta_A$ and $\beta_A^\dagger$ are the canonical antighosts of
$\gamma_A$ and $\gamma_A^\dagger$. As in the $\mathcal N=2$ case, the
constant shift does not change the gauge-fixed action, since only
$\dot\gamma_{\bar\partial}$ appears in it, but it does change the BRST
charge. After quantization the ghost brackets are
\[
 \{c,b\}=1,\qquad
 [\gamma_A,\beta_B]=[\gamma_A^\dagger,\beta_B^\dagger]
 =\delta_{AB},\qquad
 \{\zeta_A,b_{R,B}\}=\delta_{AB},
\]
and the only nonvanishing brackets of the supercharges are
$\{q,q^\dagger\}=\{\bar q,\bar q^\dagger\}=H/2$.
In a polynomial representation analogous to the one of
Section~\ref{subsec:n2-state-space}, the operators act as
\[
 \begin{gathered}
 \gamma_A=\partial_{\beta_A},
 \qquad
 \beta_A^\dagger=-\partial_{\gamma_A^\dagger},
 \qquad
 b=\partial_c,\\
 \psi^\dagger_k=\partial_{\psi^k},
 \qquad
 \bar\psi^\dagger_{\bar k}=\partial_{\bar\psi^{\bar k}},
 \qquad
 P_k=-i\partial_{X^k},
 \qquad
 \bar P_{\bar k}=-i\partial_{\bar X^{\bar k}},
 \end{gathered}
\]
on the state space
\[
 \mathcal H_{(0)}^{\mathcal N=4_{\mathrm c}}
 =\Omega^{\bullet,\bullet}(M)\otimes\Lambda\langle c\rangle
 \otimes\mathbb C[
   \beta_\partial,\beta_{\bar\partial},
   \gamma_\partial^\dagger,\gamma_{\bar\partial}^\dagger]\,
 \lvert0\rangle_0,
\]
where, similarly to the $\mathcal N=2$ case, functions of $X$ and the fermions
$\psi^k,\bar\psi^{\bar k}$ are identified with $(p,r)$-forms.

The BRST charge without R gauging is
\begin{equation}
 Q^{\mathcal N=4_{\mathrm c}}_{\mathrm{BRST}}
 =-cH
 +q\frac{\partial}{\partial\beta_\partial}
 +\gamma_\partial^\dagger q^\dagger
 +\bar q\frac{\partial}{\partial\beta_{\bar\partial}}
 +\gamma_{\bar\partial}^\dagger\bar q^\dagger
 +\frac12\sum_A\gamma_A^\dagger
       \frac{\partial}{\partial\beta_A}
       \frac{\partial}{\partial c}.
 \label{eq:n4-Qbrst-picture0}
\end{equation}
Including the superghost contributions, the two R charges become
\begin{equation}
 \begin{aligned}
 R_\partial
 &=j+\beta_\partial\gamma_\partial
   -\gamma_\partial^\dagger\beta_\partial^\dagger
 =N_\psi+N_{\beta_\partial}+N_{\gamma_\partial^\dagger},\\
 R_{\bar\partial}
 &=\bar j+\beta_{\bar\partial}\gamma_{\bar\partial}
   -\gamma_{\bar\partial}^\dagger\beta_{\bar\partial}^\dagger
 =N_{\bar\psi}+N_{\beta_{\bar\partial}}
   +N_{\gamma_{\bar\partial}^\dagger},
 \end{aligned}
 \label{eq:n4-r-charges}
\end{equation}
where $N_\psi=\psi^k\partial/\partial\psi^k$,
$N_{\bar\psi}=\bar\psi^{\bar k}\partial/\partial\bar\psi^{\bar k}$, and
$N_x$ counts powers of $x$ as in Section~\ref{subsec:n2-state-space}.
Thus $R_\partial$ counts the holomorphic fermions $\psi^k$ together with
the superghosts $\beta_\partial$ and $\gamma_\partial^\dagger$ in a
state, and $R_{\bar\partial}$ counts the antiholomorphic fermions
$\bar\psi^{\bar k}$ together with $\beta_{\bar\partial}$ and
$\gamma_{\bar\partial}^\dagger$. Gauging both R symmetries adds
$\eta R_\partial+\zeta R_{\bar\partial}$ to the BRST charge, see
Eq.~\eqref{eq:n4-quantum-gauged-twist} below. With multiplication
operators to the left of derivatives, as in
Section~\ref{subsec:n2-state-space}, both R constraints are imposed at
level zero. Section~\ref{subsec:n4-antiholomorphic-cone} imposes the
shifted constraint $T$ of Eq.~\eqref{eq:n4-quantum-gauged-twist}
absolutely and $R_\partial$ relatively. In the R-gauged model the states
also depend on the R ghosts, and the charges act on
\[
 \widehat{\mathcal H}_{(0)}^{\mathcal N=4_{\mathrm c}}
 =\Omega^{\bullet,\bullet}(M)\otimes\Lambda\langle c,\eta,\zeta\rangle
 \otimes\mathbb C[
   \beta_\partial,\beta_{\bar\partial},
   \gamma_\partial^\dagger,\gamma_{\bar\partial}^\dagger]\,
 \lvert0\rangle_0,
 \qquad b_{R,A}=\partial_{\zeta_A}.
\]

The nonvanishing ghost background replaces $\gamma_{\bar\partial}$ by
$\gamma_{\bar\partial}+u$. This adds $uD$ to the BRST charge, with
\[
 D=\bar q+\frac12\gamma_{\bar\partial}^\dagger b,
\]
and $u\beta_{\bar\partial}$ to the antiholomorphic R charge. The
twisted BRST charge and the untwisted and twisted R-gauged BRST charges
are therefore
\begin{equation}
 \begin{aligned}
 Q^{\mathcal N=4_{\mathrm c}}_{\mathrm{tw}}
 &=Q^{\mathcal N=4_{\mathrm c}}_{\mathrm{BRST}}+uD,\\
 Q^{\mathcal N=4_{\mathrm c}}_R
 &=Q^{\mathcal N=4_{\mathrm c}}_{\mathrm{BRST}}
   +\eta R_\partial+\zeta R_{\bar\partial},\\
 Q^{\mathcal N=4_{\mathrm c}}_{\mathrm{tw},R}
 &=Q^{\mathcal N=4_{\mathrm c}}_{\mathrm{BRST}}+uD+\eta R_\partial+\zeta T,
 \qquad
 T=R_{\bar\partial}+u\beta_{\bar\partial},
 \end{aligned}
 \label{eq:n4-quantum-gauged-twist}
\end{equation}
with $Q^{\mathcal N=4_{\mathrm c}}_{\mathrm{BRST}}$ given in
Eq.~\eqref{eq:n4-Qbrst-picture0}. In the rest of this section we drop
the superscript. The canonical algebra gives
\begin{equation}
 \begin{gathered}
 Q_{\mathrm{BRST}}^2=D^2=\{Q_{\mathrm{BRST}},D\}=0,\qquad [R_A,Q_{\mathrm{BRST}}]=0,\\
 [R_\partial,D]=0,\qquad [R_{\bar\partial},D]=D,\qquad
 [Q_{\mathrm{BRST}},\beta_{\bar\partial}]=D,\qquad
 [D,\beta_{\bar\partial}]=0,
 \end{gathered}
 \label{eq:n4-quantum-twist-identities}
\end{equation}
which imply that $T$ commutes with $Q_{\mathrm{tw}}$ and that $Q_R$
and $Q_{\mathrm{tw},R}$ are nilpotent. We grade the twisted complex by
$\operatorname{gh}_{\mathrm{tw}}=\operatorname{gh}+R_\partial+R_{\bar\partial}$.
The fermions and the ghosts $c,\zeta_A$ then have degree one, $\gamma_A^\dagger$
has degree two and $\beta_A$ degree zero. The background $u$ has
degree zero, so the twisted BRST charge has degree one.

\subsection{Reduction to the Dolbeault complex}
\label{subsec:n4-dolbeault-reduction}
We now compute the cohomology of the twisted BRST charges and show that the R prescriptions specified below reduce it to Dolbeault cohomology. The argument uses the polynomial state space and reduction method established for the $\mathcal N=2$ model in Section~\ref{subsec:n2-polynomial-reduction}. The ungauged calculation requires an additional change of representatives to handle the holomorphic terms.

\subsubsection{Cohomology of \texorpdfstring{$Q_{\mathrm{tw}}$ on $\mathcal H_{(0)}^{\mathcal N=4_{\mathrm c}}$}{Q\_tw for N=4c}}
\label{subsec:n4-noR-cohomology}
Consider first the part $D=\bar q+\frac12\gamma_{\bar\partial}^\dagger b$
of the twisted charge $Q_{\mathrm{tw}}$ of
Eq.~\eqref{eq:n4-quantum-gauged-twist} that is linear in $u$. On forms,
$\bar q=\bar\psi^{\bar k}\bar P_{\bar k}$ acts as $-i\bar\partial$. The
ghost term $\tfrac12\gamma_{\bar\partial}^\dagger b$ maps each state
containing $c$ to the state without $c$ and with one more power of
$\gamma_{\bar\partial}^\dagger$, exactly as in
Section~\ref{subsec:n2-noR-cohomology}. Its cohomology on
$\Lambda\langle c\rangle\otimes\mathbb C[\gamma_{\bar\partial}^\dagger]$
therefore consists of the constants. The remaining superghosts
$\gamma_\partial^\dagger$, $\beta_\partial$ and $\beta_{\bar\partial}$
do not appear in $D$, and $\bar\partial$ does not change the
holomorphic degree $p$. By the K\"unneth theorem, the cohomology of $D$
is therefore
\begin{equation}
 H(\mathcal H_{(0)}^{\mathcal N=4_{\mathrm c}},D)
 \cong\bigoplus_p H_{\bar\partial}^{p,\bullet}(M)
 \otimes\mathbb C[
 \gamma_\partial^\dagger,\beta_\partial,\beta_{\bar\partial}],
 \label{eq:n4-coho}
\end{equation}
that is, one copy of the Dolbeault cohomology in every holomorphic
degree $p$ for every monomial in the three free superghosts.

The full twisted charge $Q_{\mathrm{tw}}=Q_{\mathrm{BRST}}+uD$ does
not remove all of these copies. At fixed $u\ne0$, write
$A_u=u+\partial_{\beta_{\bar\partial}}$. Then $Q_{\mathrm{tw}}$
separates into
\[
 \begin{aligned}
 Q_{\mathrm{tw}}
 &=\underbrace{A_uD-cH+\gamma_{\bar\partial}^\dagger\bar q^\dagger}
   _{\text{same structure as }\mathcal N=2}\\
 &\quad+\underbrace{
 q\partial_{\beta_\partial}
 +\gamma_\partial^\dagger q^\dagger
 +\frac12\gamma_\partial^\dagger\partial_{\beta_\partial}b}
 _{\text{additional holomorphic terms}}.
 \end{aligned}
\]

The reduction proceeds in three steps. The change of representatives used in Section~\ref{subsec:n2-noR-cohomology} removes $-cH+\gamma_{\bar\partial}^\dagger\bar q^\dagger$ and modifies the holomorphic terms. A second conjugation removes the resulting terms proportional to $\gamma_\partial^\dagger$, and a final rescaling extracts a common factor of $A_u$. The remaining ghost calculation then leaves a differential on forms with polynomial dependence on $\beta_\partial$.

For the first step, $A_u$ is invertible on polynomials, as in Eq.~\eqref{eq:n2-Au-inverse}. Define $L=2c\bar q^\dagger A_u^{-1}$. The identity
$[D,2c\bar q^\dagger]=-cH+\gamma_{\bar\partial}^\dagger\bar q^\dagger$
gives $[A_uD,L]=-cH+\gamma_{\bar\partial}^\dagger\bar q^\dagger$. Since $L^2=0$, the transformation $1+L$ has inverse $1-L$. Conjugating $Q_{\mathrm{tw}}$ with $1+L$ gives a new differential
\[
 \begin{aligned}
 Q_1&:=(1+L)Q_{\mathrm{tw}}(1-L)\\
 &=\underbrace{A_uD+q\partial_{\beta_\partial}}_{:=Q_0}
 +\gamma_\partial^\dagger\left(
 q^\dagger-A_u^{-1}\bar q^\dagger\partial_{\beta_\partial}
 +\frac12\partial_{\beta_\partial}b\right).
 \end{aligned}
\]

The term $-\gamma_\partial^\dagger A_u^{-1}\bar q^\dagger\partial_{\beta_\partial}$ arises from transforming the mixed ghost term $\frac12\gamma_\partial^\dagger\partial_{\beta_\partial}b$.

Since $L^2=0$, the first step was a conjugation with $1+L=e^L$, and the second step uses the same method. We seek an even operator $B$ whose commutator with $Q_0$ reproduces the remaining terms $Q_1-Q_0$, so that conjugation with $e^{-B}$ removes them. The operator
$I=h^{k\bar l}\psi_k^\dagger\bar\psi_{\bar l}^\dagger$
satisfies
\[
 [I,\bar q]=q^\dagger,
 \qquad [I,q]=-\bar q^\dagger.
\]
The terms in $Q_1-Q_0$ containing $q^\dagger$ and $\bar q^\dagger$ are generated by $\gamma_\partial^\dagger A_u^{-1}I$, since
\[
 \left[\gamma_\partial^\dagger A_u^{-1}I,Q_0\right]
 =\gamma_\partial^\dagger q^\dagger
 -\gamma_\partial^\dagger A_u^{-1}\bar q^\dagger
  \partial_{\beta_\partial}.
\]
To generate the remaining term $\frac12\gamma_\partial^\dagger\partial_{\beta_\partial}b$, we use differentiation with respect to $\gamma_{\bar\partial}^\dagger$. Only the ghost term $\frac12A_u\gamma_{\bar\partial}^\dagger b$ in $Q_0$
depends on this variable, so
\[
 \left[\partial_{\gamma_{\bar\partial}^\dagger},Q_0\right]
 =\frac12A_u b.
\]
The factor
$\gamma_\partial^\dagger A_u^{-1}\partial_{\beta_\partial}$ commutes with $Q_0$. Multiplying the derivative operator by this factor therefore gives
\[
 \left[
 \gamma_\partial^\dagger A_u^{-1}
 \partial_{\beta_\partial}
 \partial_{\gamma_{\bar\partial}^\dagger},
 Q_0
 \right]
 =\frac12\gamma_\partial^\dagger\partial_{\beta_\partial}b.
\]
These two contributions motivate the generator
\[
 B=\gamma_\partial^\dagger A_u^{-1}
 \left(I+\partial_{\beta_\partial}\partial_{\gamma_{\bar\partial}^\dagger}\right),
\]
for which
\[
 \begin{aligned}
 [B,Q_0]
 &=\gamma_\partial^\dagger\left(
 q^\dagger-A_u^{-1}\bar q^\dagger\partial_{\beta_\partial}
 +\frac12\partial_{\beta_\partial}b\right),\\
 [B,[B,Q_0]]&=0.
 \end{aligned}
\]
The first identity follows from $[I,\bar q]=q^\dagger$, $[I,q]=-\bar q^\dagger$ and $\partial_{\beta_\partial}\partial_{\gamma_{\bar\partial}^\dagger}(\frac12\gamma_{\bar\partial}^\dagger b)=\frac12\partial_{\beta_\partial}b$, using that $A_u$ and $\gamma_\partial^\dagger$ commute with $D$. The double commutator vanishes because $I$ commutes with $q^\dagger$, $\bar q^\dagger$ and $b$, while $[B,Q_0]$ is independent of $\gamma_{\bar\partial}^\dagger$.
The conjugation series therefore terminates after the first commutator,
\[
 e^BQ_0e^{-B}
 =Q_0+[B,Q_0]+\frac12[B,[B,Q_0]]+\cdots
 =Q_0+[B,Q_0]=Q_1,
\]
and the inverse transformation removes these terms,
\[
 e^{-B}Q_1e^B
 =Q_0
 =A_u\left(\bar q+\frac12\gamma_{\bar\partial}^\dagger b\right)
  +q\partial_{\beta_\partial}.\]

In $Q_0=A_u(\bar q+\frac12\gamma_{\bar\partial}^\dagger b)+q\partial_{\beta_\partial}$, only the last term lacks the overall factor $A_u$. To put it in the same form, define the linear map $\mathcal C_u\colon\mathcal H_{(0)}^{\mathcal N=4_{\mathrm c}}\to\mathcal H_{(0)}^{\mathcal N=4_{\mathrm c}}$ by
\[
 \mathcal C_u(\beta_\partial^n v)
 =\beta_\partial^n A_u^n v,
 \qquad \partial_{\beta_\partial}v=0.
\]
This map is invertible because $A_u$ is, and it satisfies
\[
 \mathcal C_u^{-1}\partial_{\beta_\partial}\mathcal C_u
 =A_u\partial_{\beta_\partial}.
\]
Since $\mathcal C_u$ commutes with $A_u$, $D$ and $q$, the combined transformation is
\[
 U=\mathcal C_u^{-1}e^{-B}(1+L),
 \qquad
 UQ_{\mathrm{tw}}U^{-1}
 =A_u\left(
 \bar q+q\partial_{\beta_\partial}+\frac12\gamma_{\bar\partial}^\dagger b
 \right).
\]
All factors and their inverses preserve the polynomial state space and the twisted degree. In particular, $e^{\pm B}$ terminates on each state because each application of $B$ removes fermions or lowers the power of $\gamma_{\bar\partial}^\dagger$.

Since $\bar q+q\partial_{\beta_\partial}+\frac12\gamma_{\bar\partial}^\dagger b$ does not involve $\beta_{\bar\partial}$, it commutes with $A_u=u+\partial_{\beta_{\bar\partial}}$. Together with the invertibility of $A_u$, this implies that
$A_u(\bar q+q\partial_{\beta_\partial}+\frac12\gamma_{\bar\partial}^\dagger b)$
and $\bar q+q\partial_{\beta_\partial}+\frac12\gamma_{\bar\partial}^\dagger b$
have the same kernel and image, as in Section~\ref{subsec:n2-noR-cohomology}. The ghost differential $\frac12\gamma_{\bar\partial}^\dagger b$ has only the constant class in $\Lambda\langle c\rangle\otimes\mathbb C[\gamma_{\bar\partial}^\dagger]$, leaving
\begin{equation}
 H^\bullet(\mathcal H_{(0)}^{\mathcal N=4_{\mathrm c}},Q_{\mathrm{tw}})
 \cong H^\bullet\left(
 \Omega^{\bullet,\bullet}(M)\otimes\mathbb C[\beta_\partial],
 \bar q+q\partial_{\beta_\partial}\right)
 \otimes\mathbb C[\beta_{\bar\partial},\gamma_\partial^\dagger],
 \qquad\text{for }u\ne0.
 \label{eq:n4-noR-full-cohomology}
\end{equation}
The holomorphic term $q\partial_{\beta_\partial}$ remains in the reduced differential, which acts on forms as $\bar q+q\partial_{\beta_\partial}=-i(\bar\partial+\partial\,\partial_{\beta_\partial})$. The variables $\beta_{\bar\partial}$ and $\gamma_\partial^\dagger$ do not appear in it and give the additional polynomial factors.

\subsubsection{Cohomology of \texorpdfstring{$Q_{\mathrm{tw},R}$ on $\widehat{\mathcal H}_{(0)}^{\mathcal N=4_{\mathrm c}}$}{Q\_tw,R for N=4c}}
\label{subsec:n4-antiholomorphic-cone}
At fixed $u\ne0$, a single copy of Dolbeault cohomology is obtained by treating the two R symmetries differently. Treating the unshifted holomorphic charge relatively selects $(0,\bullet)$-forms and removes the additional holomorphic terms in $Q_{\mathrm{tw}}$. On this relative state space, $Q_{\mathrm{tw}}$ has the same
structure as in the $\mathcal N=2$ model. Treating the shifted antiholomorphic charge absolutely removes the $\beta_{\bar\partial}$ tower by the argument of Section~\ref{subsec:n2-full-brst}, leaving the Dolbeault complex with the degree shift contributed by the surviving ghost $\zeta$.

The relative prescription for $R_\partial$ restricts the state space to
\begin{equation}
 \mathcal C_{\mathrm{rel},\partial}
 :=\ker b_{R,\partial}\cap\ker R_\partial
 \subset\widehat{\mathcal H}_{(0)}^{\mathcal N=4_{\mathrm c}}.
 \label{eq:n4-holomorphic-relative-space}
\end{equation}
The antighost $b_{R,\partial}$ is represented by $\partial_\eta$, so the condition $b_{R,\partial}\Phi=0$ means that $\Phi$ is independent of $\eta$. In the representation of Eq.~\eqref{eq:n4-r-charges}, the generator $R_\partial$ acts as $N_\psi+N_{\beta_\partial}+N_{\gamma_\partial^\dagger}$, a sum of nonnegative number operators. A state in $\ker R_\partial$ is therefore independent of $\psi^k$, $\beta_\partial$ and $\gamma_\partial^\dagger$. The charge in Eq.~\eqref{eq:n4-quantum-gauged-twist} satisfies
\begin{equation}
 [Q_{\mathrm{tw},R},R_\partial]=0,
 \qquad
 \{Q_{\mathrm{tw},R},b_{R,\partial}\}=R_\partial.
 \label{eq:n4-relative-preservation}
\end{equation}
By these identities, $Q_{\mathrm{tw},R}$ preserves the common kernel in Eq.~\eqref{eq:n4-holomorphic-relative-space}. Keeping the full dependence on $\zeta$ gives
\begin{equation}
 \mathcal C_{\mathrm{rel},\partial}
 =\Lambda\langle\zeta\rangle\otimes V,
 \qquad
 V=\Omega^{0,\bullet}(M)\otimes\Lambda\langle c\rangle
 \otimes\mathbb C[\gamma_{\bar\partial}^\dagger,\beta_{\bar\partial}].
 \label{eq:n4-relative-factorization}
\end{equation}
Here the polynomial vacuum is identified with the constant $1$, as in Section~\ref{subsec:n2-state-space}.

On this space, $q\partial_{\beta_\partial}$ and $\frac12\gamma_\partial^\dagger\partial_{\beta_\partial}b$ vanish because the states are independent of $\beta_\partial$. The term $\gamma_\partial^\dagger q^\dagger$ also vanishes because $q^\dagger$ contains $\partial_{\psi^k}$. Thus
\begin{equation}
 Q_{\mathrm{tw}}\big|_V
 =\bigl(u+\partial_{\beta_{\bar\partial}}\bigr)D
   -cH+\gamma_{\bar\partial}^\dagger\bar q^\dagger,
 \qquad
 D=\bar q+\frac12\gamma_{\bar\partial}^\dagger b.
 \label{eq:n4-relative-factorization-Q}
\end{equation}
The remaining terms $-cH+\gamma_{\bar\partial}^\dagger\bar q^\dagger=[D,2c\bar q^\dagger]$ have the same structure as in the $\mathcal N=2$ model. The relative condition has removed the three holomorphic terms
discussed above Eq.~\eqref{eq:n4-relative-factorization-Q}, so
the additional conjugation and rescaling used in
Section~\ref{subsec:n4-noR-cohomology} are unnecessary here.

The term $\eta R_\partial$ vanishes on $\mathcal C_{\mathrm{rel},\partial}$, so for $f,g\in V$ the full differential acts as
\begin{equation}
 Q_{\mathrm{tw},R}(f+\zeta g)
 =Q_{\mathrm{tw}}f+\zeta\bigl(Tf-Q_{\mathrm{tw}}g\bigr),
 \qquad T=R_{\bar\partial}+u\beta_{\bar\partial}.
 \label{eq:n4-full-cone-differential}
\end{equation}
The polynomial argument of Section~\ref{subsec:n2-full-brst} applies with $\beta_{\bar\partial}$ in the role of $\beta$. The number operators in $R_{\bar\partial}$ preserve degree in $\beta_{\bar\partial}$, whereas $u\beta_{\bar\partial}$ raises it by one. Hence $T$ is injective, and successive cancellation of leading terms gives
\[
 V=TV\oplus V_0,
 \qquad
 V_0=\ker\partial_{\beta_{\bar\partial}}
 =\Omega^{0,\bullet}(M)\otimes\Lambda\langle c\rangle
 \otimes\mathbb C[\gamma_{\bar\partial}^\dagger].
\]
Both summands are preserved by $Q_{\mathrm{tw}}$. For $TV$ this follows from $[Q_{\mathrm{tw}},T]=0$. The space $V_0$ is preserved because the operator in Eq.~\eqref{eq:n4-relative-factorization-Q} contains no multiplication by $\beta_{\bar\partial}$ and therefore maps $\beta_{\bar\partial}$-independent states to $\beta_{\bar\partial}$-independent states. The argument leading to Eq.~\eqref{eq:n2-full-brst-beta-reduction} therefore gives
\[
 H^k\bigl(\mathcal C_{\mathrm{rel},\partial},Q_{\mathrm{tw},R}\bigr)
 \cong H^{k-1}(V_0,Q_{\mathrm{tw}}).
\]
Every class thus has a representative $\zeta g_0$ with $g_0\in V_0$ and $Q_{\mathrm{tw}}g_0=0$. Two such representatives define the same class exactly when their coefficients differ by a $Q_{\mathrm{tw}}$-exact state in $V_0$.

It remains to compute the cohomology of $Q_{\mathrm{tw}}$ on
$V_0$. Here $A_u$ acts as multiplication by $u$, since
$\partial_{\beta_{\bar\partial}}$ annihilates every state.
The first change of representatives in
Section~\ref{subsec:n4-noR-cohomology} therefore gives
\[
 \begin{aligned}
 Q_{\mathrm{tw}}\big|_{V_0}
 &=-cH+u\bar q+\gamma_{\bar\partial}^\dagger\bar q^\dagger
   +\frac u2\gamma_{\bar\partial}^\dagger b\\
 &=(1-L)uD(1+L),
 \qquad L=\frac{2c\bar q^\dagger}{u},\qquad L^2=0.
 \end{aligned}
\]
Thus $1+L$ gives an invertible map from $(V_0,Q_{\mathrm{tw}})$
to $(V_0,uD)$, with inverse $1-L$. Since $u\ne0$, the
differentials $uD$ and $D$ have the same kernel and image.
The two terms of $D$ act on separate tensor factors:
$\bar q$ acts on $\Omega^{0,\bullet}(M)$, while
$\frac12\gamma_{\bar\partial}^\dagger b$ acts on
$\Lambda\langle c\rangle\otimes
\mathbb C[\gamma_{\bar\partial}^\dagger]$.
The latter has only the constant cohomology class, by the
calculation in Section~\ref{subsec:n2-noR-cohomology}.
The remaining differential is therefore
$\bar q=-i\bar\partial$ on $(0,\bullet)$-forms. Consequently,
\begin{equation}
 H^k\bigl(\mathcal C_{\mathrm{rel},\partial},Q_{\mathrm{tw},R}\bigr)
 \cong H_{\bar\partial}^{0,k-1}(M),
 \qquad\text{for }u\ne0.
 \label{eq:n4-gauged-dolbeault-complex}
\end{equation}
Applying the inverse transformation $1-L$ to a $\bar\partial$-closed form $\omega\in\Omega^{0,k-1}(M)$ and multiplying by $\zeta$ gives the representative
\begin{equation}
 \Phi_\omega=\zeta\left(1-\frac{2c\bar q^\dagger}{u}\right)\omega.
 \label{eq:n4-gauged-class-representatives}
\end{equation}
The transformation $1-L$ preserves the twisted degree, while multiplication by $\zeta$ raises it by one.

Changing either R prescription changes the cohomology. Treating $\zeta$ relatively would impose $T\Phi=0$, which has no nonzero polynomial solutions. Keeping $\zeta$ absolute and also treating $\eta$ absolutely adds a second copy. To see this, decompose the full state space into eigenspaces of $R_\partial$. Since $[Q_{\mathrm{tw},R},R_\partial]=0$, the differential preserves each eigenspace. For a state satisfying $Q_{\mathrm{tw},R}\Phi_r=0$ and $R_\partial\Phi_r=r\Phi_r$ with $r>0$, Eq.~\eqref{eq:n4-relative-preservation} gives
\[
 \Phi_r
 =Q_{\mathrm{tw},R}\left(\frac1r b_{R,\partial}\Phi_r\right).
\]
Every closed state with positive holomorphic R charge is therefore exact. Only the $R_\partial=0$ sector contributes to cohomology.
Its states are independent of $\psi^k$, $\beta_\partial$
and $\gamma_\partial^\dagger$, while $\eta$ remains
unrestricted. On this sector, the term $\eta R_\partial$
vanishes, so the differential acts trivially on the factor
$\Lambda\langle\eta\rangle$. Since $\eta$ has degree one,
\[
 H^k\bigl(\widehat{\mathcal H}_{(0)}^{\mathcal N=4_{\mathrm c}},
          Q_{\mathrm{tw},R}\bigr)
 \cong H_{\bar\partial}^{0,k-1}(M)
 \oplus H_{\bar\partial}^{0,k-2}(M),
 \qquad\text{for }u\ne0.
\]

The Dolbeault complex obtained with the relative holomorphic
prescription also has a target-space interpretation.
On a Calabi--Yau threefold, taking the forms to be
Lie-algebra-valued gives the linear BV complex of holomorphic
Chern--Simons theory, the target-space field theory of the open
B-model~\cite{Witten:1992fb}. The $(0,0)$-, $(0,1)$-, $(0,2)$- and $(0,3)$-components describe the gauge ghost, gauge field, field antifield and ghost antifield, respectively. Their coefficient fields have target-space BV ghost numbers $1$, $0$, $-1$ and $-2$. Thus the coefficient field associated with a $(0,r)$-form has ghost number $1-r$, whereas the corresponding worldline state $\zeta(1-L)\omega$ has twisted degree $r+1$. The supersymmetry-ghost twist and R gauging thus give a worldline realization of the linear field complex underlying the open B-model.

\section{Closed-like models}
\label{sec:closed-like}
The models of Section~\ref{sec:open-like} have a single flavor of
worldline fermions carrying a target-space vector index, so upon
quantization their state spaces consist of differential forms. In the
untwisted particle these describe $p$-form gauge
fields~\cite{Howe:1988ft}. Upon twisting we saw that the spectra of the
$\mathcal N=2$ and $\mathcal N=4_{\mathrm c}$ models are those of
Chern--Simons and holomorphic Chern--Simons theory, consistent with the
expected connection to the A- and B-models in the open sector, whose
target-space theories are Chern--Simons theory on a Lagrangian
submanifold and holomorphic Chern--Simons
theory~\cite{Witten:1992fb}.

In this section we construct models that contain metric-like states.
For this reason we call them \emph{closed-like}, and we expect them to
match the spectra of closed-string target-space theories. We treat the
B-model, whose target-space theory is BCOV theory. To obtain such models, we introduce a flavor
index $m=1,2$ on all fermionic fields of the $\mathcal N=2$ particle.
This doubles the worldline supersymmetry from $\mathcal N=2$ to
$\mathcal N=4_{\mathrm d}$, and states now carry two independent sets of
antisymmetrized indices, such as $\psi_1^\mu\psi_2^\nu$, which include
a two-index tensor. In its real form, $\mathcal N=4_{\mathrm d}$ is a different algebra from the $\mathcal N=4_{\mathrm c}$ of the previous section. The BRST spectrum of the
$\mathcal N=4_{\mathrm d}$ spinning particle is known to contain the graviton;
$Q_{\mathrm{BRST}}$-closedness gives the linearized Einstein equations
and $Q_{\mathrm{BRST}}$-exactness linearized diffeomorphisms, and on a
curved background the BRST charge is nilpotent only if the metric
satisfies Einstein's equations~\cite{Bonezzi:2018}. The construction
extends to the NS--NS sector of supergravity~\cite{Bonezzi:2020}.
After the twist we find that the surviving states are polyvector-valued
forms, polynomial in a degree-two ghost variable $t$, with the linearized BCOV differential $\bar\partial+t\partial_\Omega$,
as expected from the closed B-model.

Section~\ref{subsec:closed-real-n4} introduces the doubled real
$\mathcal N=4_{\mathrm d}$ particle. In Section~\ref{subsec:n8-constraints} we
introduce a compatible complex structure and gauge the split
supercharges independently, which enhances the model to
$\mathcal N=8$. Section~\ref{subsec:n8-twisted-bv} gives nonzero
backgrounds to the antiholomorphic supersymmetry ghosts of both
families (the B-type choice) and applies the polynomial reduction of
Section~\ref{sec:open-like}; the BCOV differential comes from the fully
shifted BRST charge.

\subsection{The \texorpdfstring{$\mathcal N=4_{\mathrm d}$}{N=4d} particle on a real target}
\label{subsec:closed-real-n4}
The closed-like model on the same target $M$ has the same bosonic fields
$X^\mu$, $P_\mu$ and $e$, but in the fermionic sector we now introduce
a flavor index $m=1,2$ on all fermionic fields of
Section~\ref{subsec:n2-model}: the fermions become
$\psi_m^\mu,\psi^\dagger_{\mu m}$ and the gravitini
$\chi_m,\chi^\dagger_m$. The nonvanishing Poisson brackets are
\begin{equation}
 \{X^\mu,P_\nu\}_{\mathrm{cl}}=\delta^\mu{}_\nu,\qquad
 \{\psi_m^\mu,\psi^\dagger_{\nu n}\}_{\mathrm{cl}}
       =-i\delta^\mu{}_\nu\delta_{mn}.
 \label{eq:closed-n4-real-brackets}
\end{equation}
The action now reads
\begin{equation}
 \begin{aligned}
 S_0^{\mathcal N=4_{\mathrm d}}=\int d\tau\Big[
 &P_\mu\dot X^\mu
  +i\sum_m\psi^\dagger_{\mu m}\dot\psi_m^\mu
  -\frac e2H\\
 &-\frac i2\sum_m
       \bigl(\chi^\dagger_m q_m+\chi_mq^\dagger_m\bigr)\Big],
 \end{aligned}
 \label{eq:closed-n4-real-action}
\end{equation}
where the constraints are
\begin{equation}
 H=P^2,\qquad q_m=\psi_m^\mu P_\mu,\qquad
 q^\dagger_m=\eta^{\mu\nu}\psi^\dagger_{\mu m}P_\nu .
 \label{eq:closed-n4-real-charges}
\end{equation}
The two families share the Hamiltonian constraint, and
\begin{equation}
 \{q_m,q^\dagger_n\}_{\mathrm{cl}}=-i\delta_{mn}H,\qquad
 \{q_m,q_n\}_{\mathrm{cl}}
   =\{q^\dagger_m,q^\dagger_n\}_{\mathrm{cl}}=0,\qquad
 \{H,q_m\}_{\mathrm{cl}}=\{H,q^\dagger_m\}_{\mathrm{cl}}=0.
 \label{eq:closed-n4-real-algebra}
\end{equation}
There are thus four supercharges $q_m,q_m^\dagger$.

The ghosts are those of Section~\ref{subsec:n2-model}, with a flavor
index on the supersymmetry ghosts: the fermionic ghost $c$ and its
antighost $b$ for $H$, and the bosonic ghosts $\gamma_m,\gamma^\dagger_m$
with antighosts $\beta_m,\beta^\dagger_m$ for $q_m,q^\dagger_m$.
Quantization with the conventions of Section~\ref{subsec:n2-model}
replaces the Poisson brackets by graded commutators, and the only
nonvanishing bracket of the constraints is
$\{q_m,q^\dagger_n\}=\delta_{mn}H$. Before R gauging the BRST charge is
\begin{equation}
 Q^{\mathcal N=4_{\mathrm d}}_{\mathrm{BRST}}=-cH
    +\sum_m
       (\gamma_m q_m+\gamma^\dagger_m q^\dagger_m)
    +b\sum_m\gamma_m\gamma^\dagger_m,\qquad
 \bigl(Q^{\mathcal N=4_{\mathrm d}}_{\mathrm{BRST}}\bigr)^2=0.
 \label{eq:closed-n4-real-brst}
\end{equation}

For the state space we use the polynomial representation of
Section~\ref{subsec:n2-state-space}, now with two copies of the fermions and superghosts,
\begin{equation}
 \mathcal H_{(0)}^{\mathcal N=4_{\mathrm d}}
 =C^\infty(M)\otimes
  \Lambda\langle\psi_1^\mu,\psi_2^\mu,c\rangle
  \otimes\mathbb C[\gamma^\dagger_1,\gamma^\dagger_2,\beta_1,\beta_2]\,
  \lvert0\rangle_0,
 \qquad \gamma_m=\partial_{\beta_m}.
 \label{eq:closed-n4-real-states}
\end{equation}

\subsection{Complex structure and 
\texorpdfstring{$\mathcal N=8$}{N=8}}
\label{subsec:n8-constraints}
Just as the $\mathcal N=2$ particle was enhanced to $\mathcal N=4_{\mathrm c}$, the $\mathcal N=4_{\mathrm d}$ particle can be enhanced to $\mathcal N=8$ by introducing a compatible complex structure on the target space.
Each of the two real-target charge pairs splits just as in
Section~\ref{subsec:n4-model}. As in that section, from now on $q_m$ denotes the
holomorphic charge and $\bar q_m$ the antiholomorphic one,
\begin{align}
 q_m&=\psi_m^kP_k,
 &\bar q_m&=\bar\psi_m^{\bar k}\bar P_{\bar k},
 \label{eq:n8-q}\\
 q^\dagger_m&=h^{k\bar l}\psi^\dagger_{km}\bar P_{\bar l},
 &\bar q^\dagger_m&=h^{l\bar k}\bar\psi^\dagger_{\bar k m}P_l,
 \label{eq:n8-qdagger}
\end{align}
together with $H=2h^{k\bar l}P_k\bar P_{\bar l}$. The real-target
charges of Section~\ref{subsec:closed-real-n4} are the sums of the
holomorphic and antiholomorphic parts,
\begin{equation}
 \psi_m^\mu P_\mu=q_m+\bar q_m,\qquad
 \eta^{\mu\nu}\psi^\dagger_{\mu m}P_\nu=q^\dagger_m+\bar q^\dagger_m.
 \label{eq:closed-n4-n8-charge-dictionary}
\end{equation}
The gravitini, ghosts and R fields defined below carry a label
$A\in\{\partial,\bar\partial\}$ in addition to the flavor index $m$,
indicating whether they belong to the pair
$(q_m,q_m^\dagger)$ or to $(\bar q_m,\bar q_m^\dagger)$. In complex
components the fermion
brackets~\eqref{eq:closed-n4-real-brackets} become, as in
Eq.~\eqref{eq:n4-fermion-brackets},
\[
 \{\psi_m^k,\psi^\dagger_{ln}\}_{\mathrm{cl}}
 =-i\delta^k{}_l\delta_{mn},
 \qquad
 \{\bar\psi_m^{\bar k},\bar\psi^\dagger_{\bar l n}\}_{\mathrm{cl}}
 =-i\delta^{\bar k}{}_{\bar l}\delta_{mn}.
\]
Each $q^\dagger_m$ is the adjoint of $q_m$ and $\bar q^\dagger_m$ that of $\bar q_m$, and the charges therefore
satisfy
\begin{equation}
 \{q_m,q^\dagger_n\}_{\mathrm{cl}}
 =\{\bar q_m,\bar q^\dagger_n\}_{\mathrm{cl}}
 =-\frac i2\delta_{mn}H,
 \qquad
 \{q_m,\bar q^\dagger_n\}_{\mathrm{cl}}
 =\{\bar q_m,q^\dagger_n\}_{\mathrm{cl}}=0,
 \label{eq:n8-constraint-algebra}
\end{equation}
while the brackets among $q_m,\bar q_m$, among
$q^\dagger_m,\bar q^\dagger_m$, and of $H$ with all eight charges
vanish. Adding the holomorphic and antiholomorphic brackets reproduces
the algebra~\eqref{eq:closed-n4-real-algebra} of the real-target
charges.
As in the $\mathcal N=4_{\mathrm c}$ model, we gauge the eight
supercharges independently by introducing a separate pair of gravitini
$\chi_{m,A},\chi_{m,A}^\dagger$ for each $m$ and $A$. The action of the
$\mathcal N=8$ spinning particle is
\begin{equation}
 \begin{aligned}
 S_0=\int d\tau\Big[&P_k\dot X^k+\bar P_{\bar k}\dot{\bar X}^{\bar k}
 +i\sum_m(\psi^\dagger_{k m}\dot\psi_m^k
              +\bar\psi^\dagger_{\bar k m}\dot{\bar\psi}_m^{\bar k})
 -\frac e2H\\
 &-\frac i2\sum_m
 (\chi^\dagger_{m,\partial}q_m+\chi_{m,\partial}q^\dagger_m
 +\chi^\dagger_{m,\bar\partial}\bar q_m
 +\chi_{m,\bar\partial}\bar q^\dagger_m)\Big].
 \end{aligned}
 \label{eq:n8-classical-action}
\end{equation}

\paragraph{Remark: two complex structures.}
We have split both fermion families with respect to the same complex
structure. Since the two families are independent, one could instead
split $\psi_1^\mu$ and $\psi_2^\mu$ with respect to two different
complex structures $J_1$ and $J_2$, each compatible with the metric.
On a curved target this leads to bihermitian geometry, which also
arises for sigma models with $\mathcal{N}=(2,2)$ supersymmetry~\cite{Gates:1984nk}
and is equivalent to generalized K\"ahler
geometry~\cite{Gualtieri:2010}. The models of this paper correspond to
the special case $J_1=J_2$, in which the target is K\"ahler. We do not
pursue the generalized K\"ahler case here.

\subsubsection{R symmetries, BV action and quantization}
\label{subsec:n8-minimal-BV}
As for the open-like models (Section~\ref{subsec:n2-polynomial-reduction}),
applying the ghost shift to the action~\eqref{eq:n8-classical-action}
alone would leave a cohomology with infinitely many copies of the
expected field content, one for each power of the antighosts of the
shifted ghosts. We remove them, as before, by gauging R symmetries.
For each flavor $m$, and separately for the holomorphic and
antiholomorphic fermions, the global action of the R symmetry on the
matter fields is generated by
\[
 j_m=\psi_m^k\psi^\dagger_{k m},
 \qquad
 \bar j_m=\bar\psi_m^{\bar k}\bar\psi^\dagger_{\bar k m},
\]
which counts the fermions $\psi_m^k$ or $\bar\psi_m^{\bar k}$,
respectively.\footnote{The model has a larger R symmetry, including
$U(2)\times U(2)$ rotations of the two flavors of holomorphic and
antiholomorphic fermions. We gauge only the abelian generators needed
below.} We gauge the three commuting generators
\begin{equation}
 J_{\mathrm{hol}}=j_1+j_2,\qquad
 J_1=\bar j_1,\qquad J_2=\bar j_2.
 \label{eq:n8-gauged-generators}
\end{equation}
For each generator we introduce a worldline gauge field, which acts
as a Lagrange multiplier, and a fermionic R ghost: $(a_{\mathrm{hol}},\eta)$
for $J_{\mathrm{hol}}$ and $(a_m,\zeta_m)$ for $J_m$. The gauged
classical action is
\begin{equation}
 S_{0,R}^{\mathcal N=8}
 =S_0-\int d\tau\,\Bigl(a_{\mathrm{hol}}J_{\mathrm{hol}}
   +\sum_ma_mJ_m\Bigr),
 \label{eq:n8-r-gauged-classical}
\end{equation}
with $S_0$ the action from Eq.~\eqref{eq:n8-classical-action}. Since the three
generators commute with each other, their brackets impose no further
constraints. The antiholomorphic generators $J_1$ and $J_2$ are the ones shifted by
the B-type background of Section~\ref{subsec:n8-twisted-bv}, and their
R ghosts will be treated absolutely. The holomorphic generator
$J_{\mathrm{hol}}$ is not shifted and will be imposed relatively, as
$R_\partial$ in the $\mathcal N=4_{\mathrm c}$ model. A single
holomorphic generator suffices, because its relative condition already
removes the holomorphic fermions of both families.

We introduce the fermionic ghost $c$ for the constraint $H$, the
bosonic ghosts $\gamma_{m,\partial},\gamma^\dagger_{m,\partial}$ for the
symmetries generated by $q_m,q^\dagger_m$ and
$\gamma_{m,\bar\partial},\gamma^\dagger_{m,\bar\partial}$ for those
generated by $\bar q_m,\bar q^\dagger_m$. Under $J_{\mathrm{hol}}$ the ghosts
$\gamma_{m,\partial}$ and $\gamma^\dagger_{m,\partial}$ of both families
have charges $-1$ and $+1$, under $J_m$ the ghosts
$\gamma_{m,\bar\partial}$ and $\gamma^\dagger_{m,\bar\partial}$ of family
$m$ have charges $-1$ and $+1$, and all other ghosts are neutral. With
the conventions of Section~\ref{subsec:n2-r-gauging} and the antifields
on the right, the minimal BV action is the family version of
Eq.~\eqref{eq:n4-full-bv-action},
$S_{\mathrm{BV},R}^{\mathcal N=8}
=S_{0,R}^{\mathcal N=8}+\int d\tau\sum_\Phi(s\Phi)\,\Phi^*$.
For each family we write $(a_{m,\partial},\zeta_{m,\partial})=(a_{\mathrm{hol}},\eta)$
and $(a_{m,\bar\partial},\zeta_{m,\bar\partial})=(a_m,\zeta_m)$, and
$\nabla_{\tau,mA}^\pm=\partial_\tau\mp ia_{m,A}$. The BRST variations are
\begingroup\setlength{\jot}{1pt}
\begin{equation}
 \begin{aligned}
 se&=-2\dot c-\tfrac12\textstyle\sum_{m,A}
       (\gamma_{m,A}\chi_{m,A}+\gamma_{m,A}^\dagger\chi_{m,A}^\dagger),
 &sc&=\tfrac i2\textstyle\sum_{m,A}\gamma_{m,A}\gamma_{m,A}^\dagger,\\
 sX^k&=-2c\,h^{k\bar l}\bar P_{\bar l}+\textstyle\sum_m
       (\gamma_{m,\partial}\psi_m^k+\gamma_{m,\bar\partial}^\dagger h^{k\bar l}\bar\psi^\dagger_{\bar l m}),\\
 s\bar X^{\bar k}&=-2c\,h^{l\bar k}P_l+\textstyle\sum_m
       (\gamma_{m,\bar\partial}\bar\psi_m^{\bar k}
       +\gamma_{m,\partial}^\dagger h^{l\bar k}\psi^\dagger_{lm}),\\
 s\psi_m^k&=i\gamma_{m,\partial}^\dagger h^{k\bar l}\bar P_{\bar l}
       +i\eta\psi_m^k,
 &s\bar\psi_m^{\bar k}&=i\gamma_{m,\bar\partial}^\dagger h^{l\bar k}P_l
       +i\zeta_m\bar\psi_m^{\bar k},\\
 s\psi^\dagger_{km}&=i\gamma_{m,\partial}P_k-i\eta\psi^\dagger_{km},
 &s\bar\psi^\dagger_{\bar km}&=i\gamma_{m,\bar\partial}\bar P_{\bar k}
       -i\zeta_m\bar\psi^\dagger_{\bar km},\\
 s\chi_{m,A}&=-2i\nabla_{\tau,mA}^+\gamma_{m,A}^\dagger+i\zeta_{m,A}\chi_{m,A},
 &s\chi_{m,A}^\dagger&=-2i\nabla_{\tau,mA}^-\gamma_{m,A}
       -i\zeta_{m,A}\chi_{m,A}^\dagger,\\
 s\gamma_{m,A}&=-i\zeta_{m,A}\gamma_{m,A},\quad
 s\gamma_{m,A}^\dagger=i\zeta_{m,A}\gamma_{m,A}^\dagger,
 &sa_{\mathrm{hol}}&=\dot\eta,\quad sa_m=\dot\zeta_m,
 \quad s\eta=s\zeta_m=0.
 \end{aligned}
 \label{eq:n8-full-bv-action}
\end{equation}
\endgroup
As in the $\mathcal N=4_{\mathrm c}$ model, the action is linear in the
antifields. The constraint algebra~\eqref{eq:n8-constraint-algebra}
closes off shell with constant structure coefficients, so the BRST
transformations are nilpotent off shell. Since they also leave
$S_{0,R}^{\mathcal N=8}$ invariant up to a total derivative, the BV
action satisfies the classical master equation without terms quadratic
in the antifields. The variation $sc$, quadratic
in the supersymmetry ghosts, encodes the brackets
$\{q_m,q^\dagger_m\}=\{\bar q_m,\bar q^\dagger_m\}\propto H$.
Identifying
$\chi_{m,\partial}^\dagger=\chi_{m,\bar\partial}^\dagger
=\chi^\dagger_m$ and
$\chi_{m,\partial}=\chi_{m,\bar\partial}=\chi_m$
recovers Eq.~\eqref{eq:closed-n4-real-action}.

Gauge fixing proceeds as in Section~\ref{subsec:n4-gauge-fixing}. It
introduces the antighosts $b$, $\beta_{m,A}$, $\beta^\dagger_{m,A}$ and
$b_{R,\mathrm{hol}}$, $b_{R,m}$, canonically conjugate to $c$,
$\gamma_{m,A}$, $\gamma^\dagger_{m,A}$ and to the R ghosts $\eta$,
$\zeta_m$. The quantum algebra is
$\{q_m,q_n^\dagger\}=\{\bar q_m,\bar q_n^\dagger\}=\delta_{mn}H/2$, and
all other (anti)commutators of the charges vanish. The BRST charge
without R gauging is
\begin{equation}
 Q^{\mathcal N=8}_{\mathrm{BRST}}=-cH+\sum_m
 (\gamma_{m,\partial}q_m+\gamma^\dagger_{m,\partial}q^\dagger_m
 +\gamma_{m,\bar\partial}\bar q_m
 +\gamma^\dagger_{m,\bar\partial}\bar q^\dagger_m)
 +\frac12b\sum_{m,A}\gamma_{m,A}\gamma^\dagger_{m,A}.
 \label{eq:n8-operator-brst-charge}
\end{equation}
Setting both $\gamma_{m,A}=\gamma_m$ and both
$\gamma^\dagger_{m,A}=\gamma^\dagger_m$  recovers the cubic term of
Eq.~\eqref{eq:closed-n4-real-brst}. Together with the identification
of the gravitini above, these are normalization checks on the gauge
generators, not equivalences of the independently gauged complexes.
In the polynomial representation of
Section~\ref{subsec:n2-state-space} the operators are realized by the
differential operators
\begin{equation}
 \gamma_{m,A}=\partial_{\beta_{m,A}},\qquad
 \beta^\dagger_{m,A}=-\partial_{\gamma^\dagger_{m,A}},\qquad
 b=\partial_c,\qquad b_{R,\mathrm{hol}}=\partial_\eta,\qquad
 b_{R,m}=\partial_{\zeta_m}.
 \label{eq:n8-picture-zero-ghost-realization}
\end{equation}
The state space of the model without R gauging is denoted $\mathcal H_{(0)}^{\mathcal N=8}$, and that of the R-gauged model,
$\widehat{\mathcal H}_{(0)}^{\mathcal N=8}$, contains in addition the
R ghosts:
\begin{equation}
 \begin{aligned}
 \mathcal H_{(0)}^{\mathcal N=8}
 &=C^\infty(M)\otimes
 \Lambda\langle\psi_m^k,\bar\psi_m^{\bar k},c\rangle
 \otimes\mathbb C[\beta_{m,A},\gamma^\dagger_{m,A}]\,\lvert0\rangle_0,\\
 \widehat{\mathcal H}_{(0)}^{\mathcal N=8}
 &=C^\infty(M)\otimes
 \Lambda\langle\psi_m^k,\bar\psi_m^{\bar k},c,\eta,\zeta_1,\zeta_2\rangle
 \otimes\mathbb C[\beta_{m,A},\gamma^\dagger_{m,A}]\,\lvert0\rangle_0.
 \end{aligned}
 \label{eq:n8-picture-zero-state-space}
\end{equation}
Including the superghost contributions, the R charges of each family,
for the holomorphic and antiholomorphic sectors, are
\begin{equation}
 \begin{aligned}
 R_{m,\partial}
 &=j_m+\beta_{m,\partial}\gamma_{m,\partial}
              -\gamma^\dagger_{m,\partial}\beta^\dagger_{m,\partial}
 =N_{\psi_m}+N_{\beta_{m,\partial}}+N_{\gamma^\dagger_{m,\partial}},\\
 R_{m,\bar\partial}
 &=\bar j_m+\beta_{m,\bar\partial}\gamma_{m,\bar\partial}
              -\gamma^\dagger_{m,\bar\partial}\beta^\dagger_{m,\bar\partial}
 =N_{\bar\psi_m}+N_{\beta_{m,\bar\partial}}
   +N_{\gamma^\dagger_{m,\bar\partial}},
 \end{aligned}
 \label{eq:n8-r-charge}
\end{equation}
where, as in Eq.~\eqref{eq:n4-r-charges}, $N_x$ counts the powers of
$x$ in a state, and $N_{\psi_m}$, $N_{\bar\psi_m}$ count the fermions
$\psi_m^k$ and $\bar\psi_m^{\bar k}$. The full charges of the three
gauged generators~\eqref{eq:n8-gauged-generators}, with R ghosts
$\eta$, $\zeta_1$, $\zeta_2$, are
\begin{equation}
 R_{\mathrm{hol}}=R_{1,\partial}+R_{2,\partial},
 \qquad
 R_m=R_{m,\bar\partial}.
 \label{eq:n8-gauged-r-charges}
\end{equation}
With this operator ordering the R constraints are imposed at level
zero, as in Section~\ref{subsec:n2-state-space}. The R-gauged BRST
charge is
\begin{equation}
 Q^{\mathcal N=8}_R=Q^{\mathcal N=8}_{\mathrm{BRST}}+\eta R_{\mathrm{hol}}
             +\sum_m\zeta_m R_m ,
 \label{eq:n8-r-quantum-brst}
\end{equation}
which is nilpotent because the R charges commute with
$Q_{\mathrm{BRST}}$ and with each other.

We now fix how the three R ghosts are treated, using the relative and
absolute prescriptions defined in Section~\ref{subsec:n2-state-space}.
As for $R_\partial$ in the $\mathcal N=4_{\mathrm c}$ model, the
unshifted holomorphic generator is imposed relatively. We restrict to
states with $\partial_\eta\Phi=0$ and $R_{\mathrm{hol}}\Phi=0$. The first
condition says that the states do not depend on $\eta$. The second
removes all holomorphic excitations, because $R_{\mathrm{hol}}$ is a
sum of the counting operators $N_{\psi_m}$, $N_{\beta_{m,\partial}}$ and
$N_{\gamma^\dagger_{m,\partial}}$, which are nonnegative, so it vanishes
only on states containing none of $\psi_m^k$, $\beta_{m,\partial}$,
$\gamma^\dagger_{m,\partial}$. The antiholomorphic R ghosts
$\zeta_1,\zeta_2$ are treated absolutely. After the ghost shift of
Section~\ref{subsec:n8-twisted-bv}, the corresponding shifted R
constraints have vanishing kernel on polynomial states, so the
relative prescription would leave no states at all. The absolute
prescription instead removes the dependence on the antighosts conjugate to the shifted ghosts,
exactly as in Section~\ref{subsec:n2-full-brst}. These choices define
the state complex used in the following reduction.

\subsection{The B-type ghost background}
\label{subsec:n8-twisted-bv}
We now turn to the B-type twist of the $\mathcal N=8$ particle.
As in the $\mathcal N=4_{\mathrm c}$ model, the twist is induced
by a constant antiholomorphic supersymmetry-ghost background,
now involving both fermion families. The two shifted R
constraints allow the polynomial reduction of
Section~\ref{subsec:n2-full-brst} to be applied successively,
removing both $\beta$ towers. The remaining ghost calculation
leaves two matter charges, which will be identified with the
Dolbeault operator and the holomorphic divergence. A surviving
degree-two ghost multiplies the latter, giving the BCOV
coefficient differential. We choose the background by shifting
\begin{equation}
 \gamma_{m,\bar\partial}\longmapsto
        \gamma_{m,\bar\partial}+\frac{u}{\sqrt2},\qquad m=1,2,
 \label{eq:n8-background-shift}
\end{equation}
with $u$ a nonzero constant and all other ghost backgrounds zero. This selects the combination $\bar q_+$ of the rotated charges
\begin{equation}
 \bar q_\pm=\frac{\bar q_1\pm \bar q_2}{\sqrt2},\qquad
 \bar q_\pm^\dagger
 =\frac{\bar q^\dagger_1\pm \bar q^\dagger_2}{\sqrt2},
 \label{eq:n8-rotated-supercharges}
\end{equation}
which squares to zero because $\bar q_1$ and $\bar q_2$ each square to zero and anticommute.
As before, the background together with constant $X$, $e=1$ and
vanishing momenta, fermions, gravitini, R connections and antifields
solves the BV equations of motion, and the shifted action is
$S^{\mathcal N=8}_{\mathrm{tw},R}=S^{\mathcal N=8}_{\mathrm{BV},R}+uS^{\mathcal N=8}_{\mathrm{top},R}$
where
\begin{equation}
 S^{\mathcal N=8}_{\mathrm{top},R}=\frac1{\sqrt2}\int d\tau\sum_m\Big(
 \bar\psi_m^{\bar k}\bar X_{\bar k}^*
 +i\bar P_{\bar k}\bar\psi^{\dagger *\bar k}{}_m
 -\frac12\chi_{m,\bar\partial}e^*
 +\frac i2\gamma^\dagger_{m,\bar\partial}c^*
 +2a_m\chi^{\dagger *}_{m,\bar\partial}
 -i\zeta_m\gamma^*_{m,\bar\partial}\Big).
 \label{eq:n8-twisting-functional}
\end{equation}
Gauge fixing as in Section~\ref{subsec:n4-gauge-fixing} leaves the
kinetic terms unchanged. The ghost shift adds $uD$ to the BRST
charge and changes each antiholomorphic R charge by a multiple of
the antighost conjugate to the shifted ghost,
\begin{equation}
 D=\bar q_++\frac12\rho b,\qquad
 T_m=R_m+\frac{u}{\sqrt2}\beta_{m,\bar\partial},
 \qquad\text{with}\quad
 \rho=\frac{\gamma^\dagger_{1,\bar\partial}
                 +\gamma^\dagger_{2,\bar\partial}}{\sqrt2},\quad
 t=\frac{\gamma^\dagger_{1,\bar\partial}
                 -\gamma^\dagger_{2,\bar\partial}}{\sqrt2}.
 \label{eq:n8-rho-and-t}
\end{equation}
The twisted charges are therefore
\begin{equation}
 Q^{\mathcal N=8}_{\mathrm{tw}}=Q^{\mathcal N=8}_{\mathrm{BRST}}+uD,\qquad
 Q^{\mathcal N=8}_{\mathrm{tw},R}=Q^{\mathcal N=8}_{\mathrm{BRST}}+uD+\eta R_{\mathrm{hol}}
                              +\zeta_1T_1+\zeta_2T_2.
 \label{eq:n8-quantum-gauged-twist}
\end{equation}
With $D_m=\bar q_m+\frac12\gamma_{m,\bar\partial}^\dagger b$, so that
$D=(D_1+D_2)/\sqrt2$, the canonical algebra gives
$[R_m,D_n]=\delta_{mn}D_n$ and $[Q_{\mathrm{BRST}},\beta_{m,\bar\partial}]=D_m$.
As in Eq.~\eqref{eq:n4-quantum-twist-identities}, this implies
$[T_m,Q_{\mathrm{tw}}]=[T_1,T_2]=0$ and hence
$Q_{\mathrm{tw},R}^2=0$. In the twisted grading
$\operatorname{gh}_{\mathrm{tw}}=\operatorname{gh}+R_{\mathrm{hol}}+R_1+R_2$
the fermions and the odd ghosts $c,\eta,\zeta_m$ have degree one, the
ghosts $\gamma^\dagger_{m,A}$ degree two, and $\beta_{m,A}$ and $u$
degree zero, so $Q_{\mathrm{tw},R}$ has degree one.

\subsubsection{Removing the \texorpdfstring{$\beta$}{beta} towers}
\label{subsec:n8-gauged-reduction}
We compute the cohomology on the relative sector
$\mathcal C_R=\ker\partial_\eta\cap\ker R_{\mathrm{hol}}$
specified in Section~\ref{subsec:n8-minimal-BV}.
The holomorphic fermions and superghosts of both families are
absent on this sector, while both antiholomorphic R ghosts remain.
The state space is therefore
\begin{equation}
 \mathcal C_R=\Lambda\langle\zeta_1,\zeta_2\rangle\otimes V,
 \qquad
 V=C^\infty(M)\otimes
 \Lambda\langle\bar\psi_1^{\bar k},\bar\psi_2^{\bar k},c\rangle
 \otimes\mathbb C[
 \beta_{1,\bar\partial},\beta_{2,\bar\partial},
 \gamma^\dagger_{1,\bar\partial},\gamma^\dagger_{2,\bar\partial}].
 \label{eq:n8-relative-factorization}
\end{equation}
Using $D_m=\bar q_m+\frac12\gamma^\dagger_{m,\bar\partial}b$,
the restriction of the ungauged twisted charge to $V$ is
\[
 d:=Q_{\mathrm{tw}}\big|_V
 =-cH+\sum_{m=1}^2
 \left[
 \left(\frac{u}{\sqrt2}
       +\partial_{\beta_{m,\bar\partial}}\right)D_m
 +\gamma^\dagger_{m,\bar\partial}\bar q_m^\dagger
 \right].
\]
A general state has the form
$\Phi=f_0+\zeta_1f_1+\zeta_2f_2+\zeta_1\zeta_2f_{12}$,
with $f_0,f_1,f_2,f_{12}\in V$. The R-gauged charge acts as
\begin{equation}
 Q_{\mathrm{tw},R}\Phi=df_0
 +\zeta_1(T_1f_0-df_1)+\zeta_2(T_2f_0-df_2)
 +\zeta_1\zeta_2(df_{12}+T_1f_2-T_2f_1),
 \label{eq:n8-two-ghost-differential}
\end{equation}
the two-ghost version of Eq.~\eqref{eq:n2-shifted-constraint}. Each
$T_m$ commutes with $d$ and with the other constraint, and raises the
highest power of its own $\beta_{m,\bar\partial}$. We remove the two
towers one at a time. Write $\mathcal C_R=\Lambda\langle\zeta_1\rangle\otimes W$
with $W=\Lambda\langle\zeta_2\rangle\otimes V$, and let
$d_W=d+\zeta_2T_2$. Then $Q_{\mathrm{tw},R}=d_W+\zeta_1T_1$ has the form
of Eq.~\eqref{eq:n2-shifted-constraint}. $T_1$ is injective on $W$ and
commutes with $d_W$, and $d_W$ preserves
$W_0=\ker\partial_{\beta_{1,\bar\partial}}$, since neither $d$ nor $T_2$
multiplies by $\beta_{1,\bar\partial}$. The same polynomial division gives $W=T_1W\oplus W_0$. The argument of
Section~\ref{subsec:n2-full-brst} therefore gives
$H^\ell(\mathcal C_R,Q_{\mathrm{tw},R})\cong H^{\ell-1}(W_0,d_W)$. The states in $W_0$ are independent of
$\beta_{1,\bar\partial}$ but may still depend on
$\zeta_2$ and $\beta_{2,\bar\partial}$.
The operator $T_2$ preserves this independence and raises the
degree in $\beta_{2,\bar\partial}$ of every nonzero polynomial
by one. Applying the same reduction with $\zeta_2$ as the
remaining R ghost therefore gives
\begin{equation}
 H^\ell(\mathcal C_R,Q_{\mathrm{tw},R})\cong H^{\ell-2}(V_0,d),
 \qquad\text{for }u\ne0,
 \label{eq:n8-two-step-reduction}
\end{equation}
where $V_0$ is the subspace of states in $V$ independent of both
$\beta_{m,\bar\partial}$. Every class has a representative
$\zeta_1\zeta_2v$ with $v\in V_0$ and $dv=0$.
Since $\zeta_1\zeta_2$ is even, the differential on these
representatives acts as
\[
 Q_{\mathrm{tw},R}(\zeta_1\zeta_2v)=\zeta_1\zeta_2\,dv.
\]

Using
$\sum_m\gamma^\dagger_{m,\bar\partial}\bar q^\dagger_m
=\rho\,\bar q_+^\dagger+t\,\bar q_-^\dagger$, the differential on $V_0$ is
\begin{equation}
 d_0=d|_{V_0}=u \bar q_++\frac u2\rho b-cH+\rho \bar q_+^\dagger+t\bar q_-^\dagger.
 \label{eq:n8-reduced-original-brst}
\end{equation}
Compared with Eq.~\eqref{eq:n4-relative-factorization-Q}, the new
term in $d_0$ is $t\bar q_-^\dagger$. Set
$L=2c\bar q_+^\dagger/u$, so that $L^2=0$ and $1-L$ is the
inverse of $1+L$. The relation
$\{\bar q_+,\bar q_+^\dagger\}=H/2$ gives the same change of
representatives as before. The new term is unchanged because
$[L,t\bar q_-^\dagger]=0$. Thus
\begin{equation}
 (1+L)d_0(1-L)
 =u\bar q_++t\bar q_-^\dagger+\frac u2\rho b.
 \label{eq:n8-twisted-change-of-representatives}
\end{equation}
The operator $\frac u2\rho b=\frac u2\rho\,\partial_c$ acts only
on $\Lambda\langle c\rangle\otimes\mathbb C[\rho]$. For $u\ne0$,
its cohomology is represented by the constant polynomial, by the
same calculation as in Section~\ref{subsec:n2-noR-cohomology}.
The two matter charges act on the other variables, so computing
the cohomology in $c$ and $\rho$ gives
\begin{equation}
 (\mathcal C_R,Q_{\mathrm{tw},R})\simeq
 \left(
 \zeta_1\zeta_2\,C^\infty(M)\otimes
 \Lambda\langle\bar\psi_1^{\bar k},\bar\psi_2^{\bar k}\rangle[t],
 \ u\bar q_++t\bar q_-^\dagger
 \right),
 \qquad
 \Phi_\alpha=\zeta_1\zeta_2
 \left(1-\frac{2c\bar q_+^\dagger}{u}\right)\alpha.
 \label{eq:n8-gauged-matter-complex}
\end{equation}
Here $\Phi_\alpha$ represents the class of a coefficient
$\alpha$ in the displayed matter space satisfying
$(u\bar q_++t\bar q_-^\dagger)\alpha=0$.

In contrast to the open-like models, a superghost survives.
It is the degree-two combination $t$, which multiplies the
second matter charge $\bar q_-^\dagger$. This term comes from
the $u$-independent part $Q_{\mathrm{BRST}}$ of the charge.
The part linear in $u$,
$\partial_uQ_{\mathrm{tw},R}
 =D+\frac1{\sqrt2}\sum_m\zeta_m\beta_{m,\bar\partial}$,
would by itself give only $\bar q_+=-i\bar\partial$, with $t$
as a spectator.

Both nonzero shifts are needed for this reduction with the
chosen R generators. If one shift vanished, the corresponding
constraint would be $R_m$. Its nonnegative spectrum and the
identity
$\{Q_{\mathrm{tw},R},\partial_{\zeta_m}\}=R_m$
imply that only its zero-occupation sector contributes to
cohomology. This removes the corresponding fermion family and
superghosts, while its absolute R ghost remains.
Treating $\eta$ absolutely would instead add a second copy
shifted by one degree, by the argument following
Eq.~\eqref{eq:n4-gauged-class-representatives}, since
$R_{\mathrm{hol}}$ is unshifted and
$\{Q_{\mathrm{tw},R},\partial_\eta\}=R_{\mathrm{hol}}$.

\subsubsection{The BCOV coefficient complex}
\label{subsec:n8-bcov-interpretation}
After factoring out $\zeta_1\zeta_2$, the remaining task is to
identify the coefficient differential
$u\bar q_++t\bar q_-^\dagger$ with the BCOV differential.
The charge $\bar q_+$ contains the sum of the two
antiholomorphic fermions, while $\bar q_-^\dagger$
contains derivatives with respect to their difference. This suggests using the sum
for antiholomorphic forms and identifying the difference with
holomorphic polyvectors. Accordingly, we use the same sum and
difference combinations as for the charges in
Eq.~\eqref{eq:n8-rotated-supercharges},
\begin{equation}
 \bar\psi_+^{\bar k}=\frac{\bar\psi_1^{\bar k}+\bar\psi_2^{\bar k}}{\sqrt2},\qquad
 \bar\psi_-^{\bar k}=\frac{\bar\psi_1^{\bar k}-\bar\psi_2^{\bar k}}{\sqrt2},
 \qquad
 \theta_k=h_{k\bar l}\,\bar\psi_-^{\bar l},
 \label{eq:n8-rotated-fermions}
\end{equation}
and represent the new variables on the target as
\begin{equation}
 \bar\psi_+^{\bar k}\longleftrightarrow d\bar X^{\bar k},\qquad
 \theta_k\longleftrightarrow \frac{\partial}{\partial X^k}.
 \label{eq:n8-fermion-polyvector-dictionary}
\end{equation}
A function of $X$, $\bar\psi_+$ and $\theta$ with $a$ factors of
$\theta$ and $b$ factors of $\bar\psi_+$ is then a polyvector-valued
form in
$\operatorname{PV}^{a,b}(M)=\Omega^{0,b}(M,\wedge^aT^{1,0}M)$, so the
matter part of the states in Eq.~\eqref{eq:n8-gauged-matter-complex} is
$\operatorname{PV}(M)=\bigoplus_{a,b}\operatorname{PV}^{a,b}(M)$.

To define the divergence, choose a holomorphic volume form
$\Omega$. On the flat target we take
$\Omega=dX^1\wedge\dots\wedge dX^n$, where $n=\dim_{\mathbb C}M$.
We use the contraction map 
\[
 \kappa_\Omega
 \bigl(\omega\otimes(v_1\wedge\cdots\wedge v_a)\bigr)
 =\omega\wedge\iota_{v_1}\cdots\iota_{v_a}\Omega,
 \qquad \omega\in\Omega^{0,b}(M).
\]
It identifies
$\operatorname{PV}^{a,b}(M)$ with $\Omega^{n-a,b}(M)$.
The holomorphic de~Rham differential then defines
\[
 \partial_\Omega:=\kappa_\Omega^{-1}\partial\kappa_\Omega
 \colon\operatorname{PV}^{a,b}(M)
 \longrightarrow\operatorname{PV}^{a-1,b}(M).
\]
For constant $\Omega$, using left derivatives in the odd
variables, this gives
$\partial_\Omega=\partial_k\,\partial/\partial\theta_k$. In the
variables~\eqref{eq:n8-rotated-fermions} the two charges of
Eq.~\eqref{eq:n8-gauged-matter-complex} become
$\bar q_+=\bar\psi_+^{\bar k}\bar P_{\bar k}$ and
$\bar q_-^\dagger=P_k\,\partial/\partial\theta_k$, so
\begin{equation}
 \bar q_+=-i\bar\partial,\qquad \bar q_-^\dagger=-i\partial_\Omega,
 \qquad u \bar q_++t\bar q_-^\dagger=-iu\Bigl(\bar\partial+\frac tu\,\partial_\Omega\Bigr).
 \label{eq:n8-supercharges-as-bcov-operators}
\end{equation}
To make the rescaling explicit, write $t_{\mathrm{wl}}$ for the
ghost combination called $t$ in Eq.~\eqref{eq:n8-rho-and-t},
and set $t=t_{\mathrm{wl}}/u$. In this variable, the coefficient
differential becomes $-iuQ_{\mathrm{BCOV}}$, where
\begin{equation}
 \bigl(\operatorname{PV}(M)[t],Q_{\mathrm{BCOV}}\bigr),
 \qquad
 Q_{\mathrm{BCOV}}=\bar\partial+t\partial_\Omega.
 \label{eq:n8-linear-bcov-complex}
\end{equation}
Multiplication by the nonzero scalar $-iu$ does not change
kernels or images. After factoring out $\zeta_1\zeta_2$, the
cohomology is therefore that of this coefficient complex.

For a $Q_{\mathrm{BCOV}}$-closed polynomial $\alpha(t)$, the
corresponding worldline class has the representative
\[
 \Phi_\alpha=\zeta_1\zeta_2
 \left(1-\frac{2c\bar q_+^\dagger}{u}\right)
 \alpha(t_{\mathrm{wl}}/u),
\]
where the fermion identification of
Eq.~\eqref{eq:n8-fermion-polyvector-dictionary} is understood.
A monomial $t^r\alpha$ with
$\alpha\in\operatorname{PV}^{a,b}(M)$ has coefficient degree
$a+b+2r$. The rescaling and the transformation $1-L$ preserve
this degree, while multiplication by $\zeta_1\zeta_2$ raises
it by two. Consequently,
\begin{equation}
 H^\ell(\mathcal C_R,Q_{\mathrm{tw},R})
 \cong
 H^{\ell-2}\bigl(\operatorname{PV}(M)[t],Q_{\mathrm{BCOV}}\bigr),
 \qquad\text{for }u\ne0.
 \label{eq:n8-gauged-bcov-cohomology}
\end{equation}

In the BCOV field grading of
\cite{Costello:2012cy,Costello:2016}, the coefficient complex
is shifted by two. A component $t^r\alpha$ with
$\alpha\in\operatorname{PV}^{a,b}(M)$ therefore has
cohomological degree $a+b+2r-2$ in the shifted field space.
The coefficient field associated with this component has the
opposite degree, so its target-space BV ghost number is
\begin{equation}
 \operatorname{gh}_{\mathrm{target}}=2-(a+b+2r).
 \label{eq:n8-bcov-target-ghost-number}
\end{equation}
In particular, the coefficient fields associated with the
Beltrami differentials $\operatorname{PV}^{1,1}(M)$ have
ghost number zero.

Costello and Li also complete the polynomial variable to power
series, giving $\operatorname{PV}(M)[[t]]$. This completion
in $t$ is an additional choice on the target side. It is
distinct from completing the eliminated antighosts
$\beta_{m,\bar\partial}$, which would change the worldline
cohomology (Section~\ref{subsec:n2-full-brst}).
The Poisson bracket and the interactions of BCOV theory are
further target-space data that the reduction of the BRST
differential does not provide.

\section{Curved targets}
\label{sec:deformed-algebra}
Having obtained the Dolbeault and linear BCOV complexes on flat targets in Sections~\ref{sec:open-like} and~\ref{sec:closed-like}, we now ask which curved backgrounds support the corresponding twisted worldline constructions. Coupling the particle to a curved metric replaces ordinary derivatives by covariant derivatives.\footnote{This is also true for a flat target in curvilinear coordinates. In that case, however, the commutators of covariant derivatives still vanish, and the supercharge algebra is unchanged.} Their commutators introduce curvature terms into the supercharge algebra, so closure of the constraints and nilpotency of the BRST charge must be checked again. These consistency conditions restrict the admissible background geometry. The state complex is then computed with the background metric held fixed.

The open and closed sectors lead to different conditions. In the open sector, the curvature contribution can be incorporated into the Hamiltonian while preserving the full four-charge algebra on any K\"ahler target. The reduction to the Dolbeault complex therefore continues to apply. In the closed sector, we construct the curved twisted charge on the relative state space used in Section~\ref{subsec:n8-gauged-reduction}. Within the class of K\"ahler metrics, this charge is nilpotent exactly when the metric is Ricci-flat. If the target also admits a global parallel holomorphic volume form, the resulting complex is identified with the linear BCOV complex on that background.

\paragraph{Remark on target-space backgrounds.}
The reduced complexes can also be deformed by target-space background
fields. On forms with values in a Chan--Paton bundle $W$, replacing
$\bar\partial$ by $\bar\partial+A$ with
$A\in\Omega^{0,1}(M,\operatorname{End}W)$ gives a nilpotent operator
exactly when $F_A^{0,2}=\bar\partial A+A\wedge A=0$. Replacing the BCOV
differential $Q_{\mathrm{BCOV}}$ by
$Q_{\mathrm{BCOV}}+[a,-]_{\mathrm{SN}}$ with a polyvector background
$a$ gives a nilpotent operator if $a$ satisfies the extended
Kodaira--Spencer equation
$Q_{\mathrm{BCOV}}a+\frac12[a,a]_{\mathrm{SN}}=0$
\cite{BarannikovKontsevich1998Frobenius,Costello:2020}. These are the
field equations of holomorphic Chern--Simons and BCOV theory
\cite{Witten:1992fb,Bershadsky:1993cx}. Linearizing them around a
solution gives the complexes found above, with the differential
replaced by its background-covariant version. Here these couplings
are imposed on the reduced complexes rather than derived from the
worldline, so we do not pursue them further. A worldline derivation of
the Kodaira--Spencer equation from deformed particle momenta is given
in \cite[Sec.~5]{Boffo:2026}.

\subsection{K\"ahler targets in the open sector}
\label{sec:curved-n4-hcs}
The Dolbeault complex obtained in Section~\ref{subsec:n4-antiholomorphic-cone} is defined on any complex manifold. It is therefore natural to ask which curved targets admit a worldline realization of this complex. For a K\"ahler metric, we will show that a curvature coupling in the Hamiltonian preserves the four-charge algebra. We first derive this coupling at the classical level, then quantize the charges on differential forms and apply the cohomology reduction established in Section~\ref{subsec:n4-antiholomorphic-cone}.

Let $h$ be a K\"ahler metric on $M$, with Levi--Civita connection
$\Gamma$. The fermions are sections of the pullback bundle $X^*TM$, so their derivatives
along the worldline must be covariantized to transform properly under target-space diffeomorphisms:
\[
 D_\tau\psi^\lambda
 =\dot\psi^\lambda
 +\Gamma^\lambda{}_{\mu\nu}\dot X^\mu\psi^\nu.
\]
Greek indices run over both holomorphic and antiholomorphic components.
Rewriting the canonical kinetic terms using this derivative gives
\[
 P_\mu\dot X^\mu+i\psi^\dagger_\lambda\dot\psi^\lambda
 =
 \bigl(P_\mu-i\psi^\dagger_\lambda
 \Gamma^\lambda{}_{\mu\nu}\psi^\nu\bigr)\dot X^\mu
 +i\psi^\dagger_\lambda D_\tau\psi^\lambda.
\]
The connection correction to $P_\mu$ makes the coefficient of
$\dot X^\mu$ transform as a target-space covector. This defines the
covariant momentum
\begin{equation}
 \Pi_\mu
 =P_\mu-i\psi^\dagger_\lambda
       \Gamma^\lambda{}_{\mu\nu}\psi^\nu
 =P_\mu+i\Gamma^\lambda{}_{\mu\nu}
       \psi^\nu\psi^\dagger_\lambda.
 \label{eq:curved-n4-covariant-momentum}
\end{equation}
In these variables, the curved first-order action is
\begin{equation}
 \begin{aligned}
 S_0^h=\int d\tau\Big[
 &\Pi_\mu\dot X^\mu+i\psi^\dagger_\lambda D_\tau\psi^\lambda
 -\frac e2H_{\mathrm{cl}}\\
 &-\frac i2\bigl(
 \chi_\partial^\dagger q+\chi_\partial q^\dagger
 +\chi_{\bar\partial}^\dagger\bar q
 +\chi_{\bar\partial}\bar q^\dagger\bigr)
 \Big].
 \end{aligned}
 \label{eq:curved-n4-classical-action}
\end{equation}
Here $H_{\mathrm{cl}}$ will be determined by the charge algebra.
The canonical brackets of Section~\ref{subsec:n2-model} give
\begin{equation}
 \begin{aligned}
 \{\Pi_\mu,\psi^\nu\}_{\mathrm{cl}}
 &=\Gamma^\nu{}_{\mu\rho}\psi^\rho,
 &\{\Pi_\mu,\psi^\dagger_\nu\}_{\mathrm{cl}}
 &=-\Gamma^\rho{}_{\mu\nu}\psi^\dagger_\rho,\\
 \{\Pi_\mu,\Pi_\nu\}_{\mathrm{cl}}
 &=-iR^\lambda{}_{\rho\mu\nu}\psi^\rho\psi^\dagger_\lambda.
 \end{aligned}
 \label{eq:curved-n4-covariant-brackets}
\end{equation}
Here $R^\lambda{}_{\rho\mu\nu}$ is the Riemann curvature tensor. On a K\"ahler target the connection preserves the holomorphic splitting, and its curvature has type $(1,1)$. In our convention its nonvanishing components are then
$R^r{}_{s k\bar l}=-\partial_{\bar l}\Gamma^r{}_{ks}$ and their complex conjugates; this formula holds only in the K\"ahler case.

Writing $\Pi_\mu=(\Pi_k,\bar\Pi_{\bar k})$ for the holomorphic
and antiholomorphic components of the covariant momentum,
the four supercharges are the covariant versions of
Eqs.~\eqref{eq:n4-q-split} and~\eqref{eq:n4-qbar-split},
\begin{equation}
 \begin{aligned}
 q&=\psi^k\Pi_k,
 &q^\dagger&=h^{k\bar l}\psi^\dagger_k\bar\Pi_{\bar l},\\
 \bar q&=\bar\psi^{\bar k}\bar\Pi_{\bar k},
 &\bar q^\dagger&=h^{l\bar k}\bar\psi^\dagger_{\bar k}\Pi_l.
 \end{aligned}
 \label{eq:curved-n4-classical-charges}
\end{equation}
Since the charge brackets are covariant, it suffices to evaluate them at an arbitrary point in K\"ahler normal coordinates, where $\Gamma=0$ and the first derivatives of $h$ vanish. Derivatives of the connection need not vanish at that point, so the curvature term in Eq.~\eqref{eq:curved-n4-covariant-brackets} remains. In $\{q,q^\dagger\}_{\mathrm{cl}}$, the fermion bracket gives the kinetic term, while the momentum bracket in Eq.~\eqref{eq:curved-n4-covariant-brackets} gives the curvature coupling. Thus $\{q,q^\dagger\}_{\mathrm{cl}}=-iH_{\mathrm{cl}}/2$, with
\begin{equation}
 H_{\mathrm{cl}}
 =2h^{k\bar l}\Pi_k\bar\Pi_{\bar l}
 +2h^{r\bar l}R^\lambda{}_{\nu k\bar l}
       \psi^k\psi^\dagger_r\psi^\nu\psi^\dagger_\lambda.
 \label{eq:curved-n4-classical-hamiltonian}
\end{equation}
The K\"ahler curvature symmetries give the same Hamiltonian from $\{\bar q,\bar q^\dagger\}_{\mathrm{cl}}$. They also remove the possible curvature contribution to the mixed bracket. In the following expression the Greek indices $\lambda,\nu$ again run over both holomorphic and antiholomorphic components:
\[
 \{q,\bar q\}_{\mathrm{cl}}
 =-i\psi^k\bar\psi^{\bar l}
 R^\lambda{}_{\nu k\bar l}\psi^\nu\psi^\dagger_\lambda=0.
\]
For holomorphic $\nu$, the curvature coefficient is symmetric in $k$ and $\nu$ and contracts with antisymmetric fermions. For antiholomorphic $\nu=\bar n$, the curvature
coefficient is symmetric in $\bar l,\bar n$ and contracts
with the antisymmetric product
$\bar\psi^{\bar l}\bar\psi^{\bar n}$, so this contribution
also vanishes. The brackets $\{q,\bar q^\dagger\}_{\mathrm{cl}}$ and $\{\bar q,q^\dagger\}_{\mathrm{cl}}$ vanish because the curvature has no $(2,0)$ or $(0,2)$ component. The self-brackets vanish for the same reason. In normal coordinates $\{q,q\}_{\mathrm{cl}}$ reduces to $\psi^k\psi^l\{\Pi_k,\Pi_l\}_{\mathrm{cl}}$, which involves only the $(2,0)$ component $R^\lambda{}_{\rho kl}$ of the curvature, and similarly for the other self-brackets. Torsion-freeness enters through the existence of these normal coordinates. Together with the vanishing self-brackets and the corresponding relations for the daggered charges, this gives the classical algebra of Eq.~\eqref{eq:n4-constraint-algebra}, with $H$ replaced by $H_{\mathrm{cl}}$. The brackets of $H_{\mathrm{cl}}$ with the four charges then vanish by the graded Jacobi identity.

The two fermion-number charges $j = \psi^k\psi^\dagger_k$ and $\bar j = \bar\psi^{\bar k}\bar\psi^\dagger_{\bar k}$ also obey the same brackets as in the flat model. Gauging them adds $-\int d\tau\,(a_\partial j+a_{\bar\partial}\bar j)$ to the action in Eq.~\eqref{eq:curved-n4-classical-action}. Since the constraint algebra has the same structure constants, the multiplier and ghost transformations in Eq.~\eqref{eq:n4-full-bv-action} keep their form. The matter transformations are generated by the covariant constraints and acquire connection and curvature terms. The background of Section~\ref{subsec:n4-flat-bv-twist} remains a solution of the classical BV equations: $X$ is constant, $e=1$, $\gamma_{\bar\partial}=u$, and all other fields and antifields vanish. In particular, the additional terms vanish when the momenta and fermions vanish. The shift $\gamma_{\bar\partial}\mapsto\gamma_{\bar\partial}+u$ therefore defines the curved BV twist as before.

Upon quantization, the fermions act by exterior multiplication and their canonical partners by contraction. In this representation,
$\psi^\dagger_\lambda=\partial/\partial\psi^\lambda$ and
$P_\mu=-i\partial_\mu$. Placing the contractions on the right
in Eq.~\eqref{eq:curved-n4-covariant-momentum} makes
$\Pi_\mu$ act as $-i$ times the Levi--Civita covariant
derivative on differential forms,
\[
 \Pi_\mu=-i\nabla_\mu,
 \qquad
 \nabla_\mu=\partial_\mu
 -\Gamma^\lambda{}_{\mu\nu}\psi^\nu
       \frac{\partial}{\partial\psi^\lambda}.
\]
For example,
\[
 q=-i\psi^k\nabla_k
 =-i\psi^k\partial_k
 +i\Gamma^l{}_{km}\psi^k\psi^m
       \frac{\partial}{\partial\psi^l}
 =-i\partial,
\]
because $\Gamma^l{}_{km}$ is symmetric in $k,m$. The same argument gives $\bar q=-i\bar\partial$. Taking the daggered charges to be the formal adjoints for the Hermitian inner product fixes their ordering and yields
\begin{equation}
 q=-i\partial,\qquad
 \bar q=-i\bar\partial,\qquad
 q^\dagger=i\partial^\dagger,\qquad
 \bar q^\dagger=i\bar\partial^\dagger.
 \label{eq:curved-n4-quantum-charges}
\end{equation}
The K\"ahler identities needed for the quantum algebra are
\[
 \{\partial,\bar\partial^\dagger\}
 =\{\bar\partial,\partial^\dagger\}=0,
 \qquad
 \Delta_\partial=\Delta_{\bar\partial}=\frac12\Delta_d.
\]
Together with $\partial^2=\bar\partial^2=\{\partial,\bar\partial\}=0$ and their adjoint relations, they give
\begin{equation}
 \{q,q^\dagger\}=\{\bar q,\bar q^\dagger\}
 =\frac12H,
 \qquad H=\Delta_d,
 \label{eq:curved-n4-quantum-algebra}
\end{equation}
with all other anticommutators of the four charges vanishing. The Hamiltonian commutes with each charge. The relation $H=2\{q,q^\dagger\}$ fixes the ordering of the
quantum Hamiltonian and identifies it with $\Delta_d$. The algebra holds for every K\"ahler metric, without a Ricci-flatness condition.

Let $Q^h_{\mathrm{tw},R}$ denote the charge of Eq.~\eqref{eq:n4-quantum-gauged-twist} formed with the covariant matter operators above. At fixed $u\ne0$, we use the polynomial ghost representation and impose the holomorphic R constraint relatively, while treating the shifted antiholomorphic constraint absolutely. The resulting state space is
\[
 \mathcal C_{\mathrm{rel},\partial,h}
 =\Lambda\langle\zeta\rangle\otimes\Omega^{0,\bullet}(M)
 \otimes\Lambda\langle c\rangle
 \otimes\mathbb C[\gamma_{\bar\partial}^\dagger,\beta_{\bar\partial}].
\]
The covariant supercharges have the same holomorphic and antiholomorphic degrees as their flat counterparts, so $Q^h_{\mathrm{tw},R}$ preserves this relative state space. The R generators still act as number operators, and the ghost dependence of the BRST charge is unchanged. Together with the quantum algebra in Eq.~\eqref{eq:curved-n4-quantum-algebra}, these properties ensure that the cohomology reduction of Section~\ref{subsec:n4-antiholomorphic-cone} applies unchanged. The reduction first removes the $\beta_{\bar\partial}$ tower. After an invertible change of representatives, the remaining $c,\gamma_{\bar\partial}^\dagger$ ghost complex contributes only its constant cohomology class. The remaining differential is $-iu\bar\partial$, whose cohomology is Dolbeault cohomology for $u\ne0$. The surviving R ghost $\zeta$ raises the degree by one, giving
\begin{equation}
 H^k(\mathcal C_{\mathrm{rel},\partial,h},Q^h_{\mathrm{tw},R})
 \cong H_{\bar\partial}^{0,k-1}(M),
 \qquad u\ne0.
 \label{eq:curved-n4-gauged-dolbeault-result}
\end{equation}
Applying the inverse change of representatives to a $\bar\partial$-closed form $\omega\in\Omega^{0,k-1}(M)$ gives
\[
 \Phi_\omega
 =\zeta\left(1-\frac{2c\bar q^\dagger}{u}\right)\omega.
\]
All operators used in the reduction are globally defined and preserve compact support. With compact support imposed on the forms throughout, the same argument gives compactly supported Dolbeault cohomology.

\subsection{Ricci-flat targets in the closed sector}
\label{sec:nonlinear-bcov-reduced}

Having obtained the BCOV complex from the flat closed model, we now consider how this worldline description extends to curved targets. The starting point is the covariantization of Section~\ref{sec:curved-n4-hcs}, applied to both fermion families. Replacing the fermion derivatives in the kinetic terms by $D_\tau\psi_m^\lambda=\dot\psi_m^\lambda+\Gamma^\lambda{}_{\mu\nu}\dot X^\mu\psi_m^\nu$ gives the covariant momentum
\[
 \Pi_\mu=P_\mu+i\Gamma^\lambda{}_{\mu\nu}
              \sum_{m=1}^2\psi_m^\nu\psi^\dagger_{m,\lambda}.
\]
Upon quantization, placing the fermionic derivatives on the right gives
\[
 \Pi_\mu=-i\nabla_\mu,
 \qquad
 \nabla_\mu=\partial_\mu-\Gamma^\lambda{}_{\mu\nu}
              \sum_{m=1}^2\psi_m^\nu
              \frac{\partial}{\partial\psi_m^\lambda}.
\]
The sum implements the Levi--Civita connection on both fermion families, so $\nabla_\mu$ now acts on states carrying two sets of antisymmetric tensor indices. Greek indices include holomorphic and antiholomorphic components.

The additional family changes the closure of the supercharge algebra. With the convention $[\nabla_k,\nabla_{\bar l}]\psi_m^r=-R^r{}_{sk\bar l}\psi_m^s$, the covariant charges give, for example,
\begin{equation}
 \{q_1,\bar q_1\}\psi_2^r
 =R^r{}_{s k\bar l}\psi_1^k\bar\psi_1^{\bar l}\psi_2^s.
 \label{eq:n9-curvature-bracket}
\end{equation}
This is generally nonzero on a curved target, whereas the flat eight-charge algebra requires this anticommutator to vanish. Requiring that algebra on the full state space would therefore force the metric to be flat. The construction below instead starts on the relative sector of Eq.~\eqref{eq:n8-relative-factorization}, where the holomorphic fermions and superghosts are absent. The result derived in this section is a curved quantum complex on this sector, rather than a curved BV action for the full eight-charge model. Nilpotency of the relative charge will require Ricci-flatness. Once this condition is established, the cohomology reduction of Section~\ref{subsec:n8-gauged-reduction} will identify the remaining differential, and a parallel holomorphic volume form will relate it to the BCOV differential.

On the relative sector, only the antiholomorphic fermions remain. As in Eq.~\eqref{eq:n8-rotated-fermions}, set
\[
 \bar\psi_\pm^{\bar k}
 =\frac{\bar\psi_1^{\bar k}\pm\bar\psi_2^{\bar k}}{\sqrt2},
 \qquad
 \theta_k=h_{k\bar l}\bar\psi_-^{\bar l}.
\]
The rotation from the two original families to the $+$ and $-$ variables has constant coefficients, so it preserves the summed connection term in $\nabla_\mu$. The interpretation of the new variables as tensor basis elements is supplied by the dictionary of Eq.~\eqref{eq:n8-fermion-polyvector-dictionary},
\[
 \bar\psi_+^{\bar k}\longleftrightarrow d\bar X^{\bar k},
 \qquad
 \theta_k\longleftrightarrow\frac{\partial}{\partial X^k}.
\]
This identifies the matter states with polyvector-valued forms,
\[
 \operatorname{PV}(M)=\bigoplus_{a,b}\Omega^{0,b}(M,\wedge^aT^{1,0}M).
\]
Writing $\beta_m=\beta_{m,\bar\partial}$ and $\gamma_m^\dagger=\gamma_{m,\bar\partial}^\dagger$, the full relative state space is
\[
 \mathcal C_{R,h}
 =\Lambda\langle\zeta_1,\zeta_2\rangle\otimes V_h,
 \qquad
 V_h=\operatorname{PV}(M)\otimes\Lambda\langle c\rangle
       \otimes\mathbb C[\beta_1,\beta_2,\gamma_1^\dagger,\gamma_2^\dagger].
\]
The R generators $R_m$ still refer to the original fermion families, as in Eq.~\eqref{eq:n8-gauged-r-charges}.

To express the covariant derivative in these variables, note that the K\"ahler connection has $\Gamma^{\bar r}{}_{k\bar s}=0$. On the relative sector, $\nabla_k$ therefore acts as $\partial_k$ on the coefficient functions of a state expanded in monomials of $\bar\psi_\pm$, holding these fermionic variables fixed. The substitution $\theta_s=h_{s\bar l}\bar\psi_-^{\bar l}$ depends on the target coordinates through the metric. Expressing the same covariant derivative in the $(\bar\psi_+,\theta)$ variables therefore introduces a term from differentiating $h$. The chain rule gives
\[
 \begin{aligned}
 \nabla_k
 &=\partial_k+(\partial_k h_{s\bar l})\bar\psi_-^{\bar l}
                  \frac{\partial}{\partial\theta_s}\\
 &=\partial_k+\Gamma^r{}_{ks}\theta_r
                  \frac{\partial}{\partial\theta_s}.
 \end{aligned}
\]
Here $\partial_k$ on the right holds $\bar\psi_+$ and $\theta$ fixed, and the last equality uses metric compatibility, $\partial_k h_{s\bar l}=\Gamma^r{}_{ks}h_{r\bar l}$.

The covariant charge $\bar q_-^\dagger$ contracts $\nabla_k$ on the left with $-i\partial/\partial\theta_k$. Thus its divergence operator is $\delta_h=(\partial/\partial\theta_k)\nabla_k$. Using left derivatives in the odd variables, the symmetry of $\Gamma^r{}_{ks}$ cancels the term with two $\theta$ derivatives and leaves the trace $\Gamma^l{}_{lk}=\partial_k\log\det h$. The three charges needed for the reduction therefore act as
\begin{equation}
 \begin{gathered}
 \bar q_+=-i\bar\partial,
 \qquad\bar q_+^\dagger=i\bar\partial^\dagger,
 \qquad\bar q_-^\dagger=-i\delta_h,\\
 \delta_h\alpha
 =\partial_k\left(\frac{\partial\alpha}{\partial\theta_k}\right)
   +\bigl(\partial_k\log\det h\bigr)
      \frac{\partial\alpha}{\partial\theta_k}.
 \end{gathered}
 \label{eq:n9-surviving-charges}
\end{equation}
Here $\bar\partial^\dagger$ is the formal adjoint of $\bar\partial$ for the Hermitian inner product on polyvector-valued forms induced by $h$.

On any K\"ahler target these operators satisfy
\[
 \bar q_+^2=(\bar q_+^\dagger)^2=(\bar q_-^\dagger)^2=0,
 \qquad
 \{\bar q_+^\dagger,\bar q_-^\dagger\}=0,
 \qquad
 \{\bar q_+,\bar q_+^\dagger\}=\frac H2,
 \qquad H:=2\Delta_{\bar\partial}.
\]
The relation $\{\bar q_+^\dagger,\bar q_-^\dagger\}=0$ follows from $\bar\partial^\dagger=-h^{k\bar l}\partial_{\bar\psi_+^{\bar l}}\nabla_k$ and $\delta_h=\partial_{\theta_k}\nabla_k$, using metric compatibility and $[\nabla_k,\nabla_l]=0$. The anticommutator that will obstruct nilpotency is
\begin{equation}
 \{\bar q_+,\bar q_-^\dagger\}
 =-\bar\psi_+^{\bar l}\partial_{\bar l}\partial_k\log\det h
       \frac{\partial}{\partial\theta_k}
 =\operatorname{Ric}_{k\bar l}\bar\psi_+^{\bar l}
       \frac{\partial}{\partial\theta_k}.
 \label{eq:n9-ricci-obstruction}
\end{equation}
Here $\operatorname{Ric}_{k\bar l}=-\partial_k\partial_{\bar l}\log\det h$ are the components of the Ricci tensor. With the curvature convention above, $\operatorname{Ric}_{k\bar l}=R^r{}_{rk\bar l}$. Locally, applying the bracket to $\theta_k$ gives $\operatorname{Ric}_{k\bar l}\bar\psi_+^{\bar l}$, so it vanishes on all polyvector-valued forms if and only if $h$ is Ricci-flat.

Using $\bar q_-=(\bar q_-^\dagger)^\dagger$ and the inverse rotation of Eq.~\eqref{eq:n8-rotated-supercharges} defines the covariant charges $\bar q_m,\bar q_m^\dagger$. The candidate twisted charge on $\mathcal C_{R,h}$ is then
\begin{equation}
 \begin{aligned}
 Q^h_{\mathrm{tw},R}&=d_h+\zeta_1T_1+\zeta_2T_2,
 \qquad T_m=R_m+\frac{u}{\sqrt2}\beta_m,\\
 d_h&=-cH+\sum_{m=1}^2\left[
       \bar q_m\left(\partial_{\beta_m}+\frac{u}{\sqrt2}\right)
       +\gamma^\dagger_m\bar q_m^\dagger
       +\frac12\gamma^\dagger_m
            \left(\partial_{\beta_m}+\frac{u}{\sqrt2}\right)b
       \right].
 \end{aligned}
 \label{eq:n9-curved-r-charge}
\end{equation}
This has the form of Eq.~\eqref{eq:n8-quantum-gauged-twist}, but its nilpotency must now be checked using the curved charge algebra.

\paragraph{Ricci-flatness is necessary.}
To test nilpotency, consider the subspace $\zeta_1\zeta_2V_{0,h}$, where
\[
 V_{0,h}:=\ker\partial_{\beta_1}\cap\ker\partial_{\beta_2}
          \subset V_h.
\]
This subspace is preserved by the candidate charge on any K\"ahler target. The $\zeta_mT_m$ terms annihilate it because $\zeta_m^2=0$, and $d_h$ preserves independence of both $\beta_m$. Thus
\[
 Q^h_{\mathrm{tw},R}(\zeta_1\zeta_2v)
 =\zeta_1\zeta_2d_{0,h}v,
 \qquad
 d_{0,h}=u\bar q_++t\bar q_-^\dagger
          +\frac u2\rho b-cH+\rho\bar q_+^\dagger,
\]
with $\rho=(\gamma_1^\dagger+\gamma_2^\dagger)/\sqrt2$ and $t=(\gamma_1^\dagger-\gamma_2^\dagger)/\sqrt2$ as in Eq.~\eqref{eq:n8-rho-and-t}.

Set $L=2c\bar q_+^\dagger/u$. Since $L^2=0$, the operators $1+L$ and $1-L$ are inverse. The conjugation of Eq.~\eqref{eq:n8-twisted-change-of-representatives} uses $\{\bar q_+,\bar q_+^\dagger\}=H/2$ to remove $-cH+\rho\bar q_+^\dagger$. It leaves $t\bar q_-^\dagger$ unchanged because $\{\bar q_+^\dagger,\bar q_-^\dagger\}=0$. Both relations hold before imposing Ricci-flatness, so
\begin{equation}
 (1+L)d_{0,h}(1-L)
 =u\bar q_++t\bar q_-^\dagger+\frac u2\rho b
 =-i(u\bar\partial+t\delta_h)+\frac u2\rho b.
 \label{eq:n9-curved-similarity}
\end{equation}
The ghost term squares to zero and anticommutes with the matter terms. The square of the displayed operator is therefore $ut\{\bar q_+,\bar q_-^\dagger\}$. Nilpotency on $\mathcal C_{R,h}$ requires this square to vanish. Since $u\ne0$ and multiplication by $t$ is injective on polynomial states, Eq.~\eqref{eq:n9-ricci-obstruction} implies
\begin{equation}
 \bigl(Q^h_{\mathrm{tw},R}\bigr)^2=0
 \quad\Longrightarrow\quad
 \operatorname{Ric}(h)=0.
 \label{eq:n9-ricci-flatness}
\end{equation}

\paragraph{Ricci-flatness is sufficient.}
It remains to check nilpotency on the entire relative state space. To establish the required charge algebra, we identify the covariant charges locally with operators on differential forms and then apply the K\"ahler identities. If $\operatorname{Ric}(h)=0$, the induced connection on the canonical bundle is flat. Each simply connected coordinate neighborhood therefore admits a parallel holomorphic volume form $\Omega=f(X)\,dX^1\wedge\cdots\wedge dX^n$. The equation $\nabla\Omega=0$ gives $\partial_k\log f=\Gamma^l{}_{lk}$, so
\begin{equation}
 \delta_h=\partial_\Omega,
 \qquad
 \partial_\Omega\alpha
 =\partial_k\left(\frac{\partial\alpha}{\partial\theta_k}\right)
  +\bigl(\partial_k\log f\bigr)
       \frac{\partial\alpha}{\partial\theta_k}.
 \label{eq:n9-pv-divergence}
\end{equation}
The contraction map $\kappa_\Omega$ of Section~\ref{subsec:n8-bcov-interpretation} sends $\bar q_+$ and $\bar q_-^\dagger$ to $-i\bar\partial$ and $-i\partial$ on differential forms. A parallel $\Omega$ has constant norm, so this map preserves the metric inner products up to a constant and also intertwines formal adjoints. In particular, $\bar q_-$ becomes $i\partial^\dagger$.

The K\"ahler identities now give all the required relations. The anticommutators between the $+$ and $-$ charge pairs vanish, while $\Delta_\partial=\Delta_{\bar\partial}$ gives the same Hamiltonian for both pairs. Rotating back to the original family basis gives
\begin{equation}
 \{\bar q_m,\bar q_n^\dagger\}=\tfrac12\delta_{mn}H,
 \qquad
 \{\bar q_m,\bar q_n\}=\{\bar q_m^\dagger,\bar q_n^\dagger\}=0,
 \qquad
 [H,\bar q_m]=[H,\bar q_m^\dagger]=0.
 \label{eq:n9-active-charge-algebra}
\end{equation}
Although a local volume form was used to prove these identities, the charges themselves are globally defined, so the identities hold on all of $M$.

Equation~\eqref{eq:n9-active-charge-algebra} gives $d_h^2=0$ by the same cancellation of Hamiltonian and ghost terms as in the flat model. The covariant supercharges have the same R weights as their flat counterparts, and the common Hamiltonian is R neutral. Consequently, with $D_m:=\bar q_m+\frac12\gamma_m^\dagger b$,
\[
 [R_m,d_h]=\frac{u}{\sqrt2}D_m,
 \qquad [\beta_m,d_h]=-D_m,
 \qquad [T_m,d_h]=0.
\]
Together with $[T_1,T_2]=0$, this proves nilpotency of $Q^h_{\mathrm{tw},R}$. Combining necessity and sufficiency yields
\begin{equation}
 \bigl(Q^h_{\mathrm{tw},R}\bigr)^2=0
 \quad\Longleftrightarrow\quad
 \operatorname{Ric}(h)=0.
 \label{eq:n9-curved-r-nilpotency}
\end{equation}

\paragraph{Identification with the BCOV complex.}
For a Ricci-flat metric, the charge algebra in Eq.~\eqref{eq:n9-active-charge-algebra}, together with the unchanged polynomial ghost representation and R prescriptions, ensures that the cohomology reduction of Section~\ref{subsec:n8-gauged-reduction} applies. Combined with the change of representatives in Eq.~\eqref{eq:n9-curved-similarity}, this gives the coefficient complex
\[
 \left(\operatorname{PV}(M)[t_{\mathrm{wl}}],
       -i(u\bar\partial+t_{\mathrm{wl}}\delta_h)\right),
\]
with the surviving factor $\zeta_1\zeta_2$. Here $t_{\mathrm{wl}}=(\gamma_1^\dagger-\gamma_2^\dagger)/\sqrt2$ denotes the superghost called $t$ in Eq.~\eqref{eq:n8-rho-and-t}. As in Section~\ref{subsec:n8-bcov-interpretation}, we rename it so that $t$ can denote the BCOV variable.

The BCOV coefficient differential $\bar\partial+t\partial_\Omega$ is defined by a global nowhere-vanishing holomorphic volume form $\Omega$ and does not require a metric. The worldline reduction yields the metric-dependent divergence $\delta_h$. If $\Omega$ is parallel with respect to $h$, Eq.~\eqref{eq:n9-pv-divergence} gives $\delta_h=\partial_\Omega$ globally.
On a compact Ricci-flat K\"ahler target, a global holomorphic volume form is automatically parallel by Bochner's theorem. In general, a global parallel volume form exists exactly when the flat canonical-bundle connection has trivial holonomy, as happens, for example, when $M$ is simply connected.

Setting $t=t_{\mathrm{wl}}/u$ gives the differential $-iu(\bar\partial+t\partial_\Omega)$. Multiplying each homogeneous element of coefficient degree $k$ by $(-iu)^{-k}$ removes the overall factor by a chain isomorphism. Thus
\begin{equation}
 (\mathcal C_{R,h},Q^h_{\mathrm{tw},R})
 \simeq
 \left(\zeta_1\zeta_2\operatorname{PV}(M)[t],
       \bar\partial+t\partial_\Omega\right).
 \label{eq:n9-linear-charge}
\end{equation}
The factor $\zeta_1\zeta_2$ raises the degree by two, as in Eq.~\eqref{eq:n8-gauged-bcov-cohomology}. The metric $h$ specifies the Ricci-flat background on which the worldline operators are defined, while the reduced fields are the polyvector-valued coefficients of the linear BCOV complex.

\section{Conclusion}
\label{sec:conclusion}
In this paper we constructed a worldline analogue of the Costello--Li twist~\cite{Costello:2016} for spinning particles. The twisted BRST charges can be constructed both directly in the quantum operator algebra and derived from the classical BV action. The BV derivation introduces constant backgrounds for the bosonic ghosts of local worldline supersymmetry and reproduces the twisted charges through gauge fixing and canonical quantization. Using polynomial superghost states and the absolute and relative R-ghost prescriptions, we have shown that the resulting BRST complexes reduce to the de~Rham, Dolbeault and linear BCOV complexes. The Dolbeault and BCOV reductions thus provide a worldline realization of the linear field complexes of the open and closed B-model target-space theories, respectively.

The extension to curved targets shows how consistency of the worldline construction constrains the background geometry. In the open sector, the charge algebra and the reduction to the Dolbeault complex extend to any K\"ahler target. In the closed sector, the covariant charge on the chosen relative state space is nilpotent if and only if the reference K\"ahler metric is Ricci-flat. A global identification of its reduction with the BCOV complex additionally uses a parallel holomorphic volume form.

These results establish the comparison at the level of linear field complexes. The worldline formalism alone does not constrain the target-space interaction terms. The dependence on the superghost representation also deserves further study. Our reductions use polynomial states, while other representations contain additional sectors of the BRST spectrum~\cite{Boffo:2024,Boffo:2025pcm}. Extending the twist to higher-picture sectors would clarify how the resulting spectrum and its target-space interpretation depend on this choice.

The natural next step is to extend this construction to the full RNS worldsheet theory, following the supersymmetry-ghost background through gauge fixing and quantization. The ultimate goal is to derive the topological B-model within this framework and establish a correspondence between the worldsheet twist and the target-space twist proposed by Costello and Li.

\section*{Acknowledgments}
\addcontentsline{toc}{section}{Acknowledgments}
We thank Fabian Hahner for discussions.
The work of OH is supported by the Czech Science Foundation GA\v{C}R
through the grant ``On and Off-shell Superspace Representations''
(GA26-23375S). The work of JG is supported by the Deutsche Forschungsgemeinschaft (DFG, German
Research Foundation) under Germany’s Excellence Strategy EXC 2181/1 -- 390900948 (the
Heidelberg STRUCTURES Excellence Cluster).

\end{document}